\documentclass{aa}  

\usepackage{cuted}
\usepackage{graphicx}
\usepackage{footnote}
\usepackage{booktabs}

\makesavenoteenv{tabular}
\makesavenoteenv{table}

\usepackage{txfonts}

\newcommand{\loggf}[1]{$\log(gf)_\mathrm{#1}$}
\newcommand{\dlam}[0]{$\Delta\lambda$}

\newcommand{\logtau}[1]{$\log\tau_\mathrm{#1}$}
\newcommand{\globin}{\texttt{globin}}
\newcommand{\Globin}[1]{coupled}

\newcommand{\eq}[1]{Eq.~(\ref{#1})} 
\newcommand{\fig}[1]{Fig.~\ref{#1}} 
\newcommand{\sref}[1]{Sect.~\ref{#1}} 
\newcommand{\tab}[1]{Tab.~\ref{#1}} 

\begin{document}
   \title{\globin{}: A spectropolarimetric inversion code for the coupled inference of atomic line parameters}

   \author{D. Vukadinovi\' c\inst{1,2},
          H. N. Smitha\inst{1},
          A. Korpi-Lagg\inst{1,3},
          M. van Noort\inst{1},
          J. S. Castellanos Dur\' an\inst{1}
          and
          S. K. Solanki\inst{1}
          }

   \institute{Max Planck Institute for Solar System Research, Justus-von-Liebig Weg 3, Göttingen \\
             \email{vukadinovic@mps.mpg.de}
         \and
         Georg-August-Universität Göttingen, Friedrich-Hund-Platz 1, 37077 Göttingen, Germany
         \and
         Department of Computer Science, Aalto University, PO Box 15400, FI-00076 Aalto, Finland
             }

   \date{Received month dd, yyyy; accepted month dd, yyy}

\authorrunning{Vukadinovi\'{c} et al.}
\titlerunning{\globin{}: A spectropolarimetric inversion code for the coupled inference of atomic line parameters}

  \abstract
  {The reliability of physical parameters describing the solar atmosphere inferred from observed spectral line profiles depends on the accuracy of the involved atomic parameters. For many transitions, atomic data, such as the oscillator strength (\loggf{}) and the central wavelength of the line, are poorly constrained or even unknown.}
  {We present and test a new inversion method
  that infers atomic line parameters and the height stratification of the atmospheric parameters from spatially resolved spectropolarimetric observations of the Sun. This method is implemented in the new inversion code \globin{}.
  }
  {The new method employs a global minimization algorithm enabling the coupling of inversion parameters common to all pixels, such as the atomic parameters of the observed spectral lines. At the same time, it permits the optimum atmospheric parameters to be retrieved individually for each spatial pixel. The uniqueness of this method lies in its ability to retrieve reliable atomic parameters even for heavily blended spectral lines. We tested the method by applying it to a set of 18 blended spectral lines between 4015\,\AA{} and 4017\,\AA{}, synthesized from a 3D magnetohydrodynamic simulation containing a sunspot and the quiet Sun region around it. The results were then compared with a previously used inversion method where atomic parameters were determined for every pixel independently (pixel-by-pixel method). For the same spectral region, we also inferred the atomic parameters from the synthesised spatially averaged disc-centre spectrum of the quiet-sun.}
  {The new method was able to retrieve the \loggf{} values of all lines to an accuracy of 0.004\,dex, while the pixel-by-pixel method retrieved the same parameter to an accuracy of only 0.025\,dex. The largest differences between the two methods are evident for the heavily blended lines, with the former method performing better than the latter. In addition, the new method is also able to infer reliable atmospheric parameters in all the inverted pixels by successfully disentangling the degeneracies between the atomic and atmospheric parameters.}
  {The new method is well suited for the reliable determination of both atomic and atmospheric parameters and works well on all spectral lines, including those that are weak and/or severely blended. This is of high relevance, especially for the analysis of observations of spectral regions with a very high density of spectral lines. An example includes the future near-ultraviolet spectropolarimetric observations of the \textsc{Sunrise iii} stratospheric balloon mission.}

   \maketitle
 

\section{Introduction}

The physical conditions in the solar atmosphere are encoded in the spectral lines, arising from absorption and emission processes in atoms, ions, and molecules. By modelling the observed spectral line profiles and their polarization states, we can retrieve a height stratification of the atmospheric parameters, such as temperature, magnetic field, and line-of-sight velocity.

To model the observed line profiles, the spectral lines are synthesized first with an initial guess model for the atmosphere and then compared with the observations. The differences between the observed and the synthetic spectra are iteratively reduced by adjusting  the input atmospheric model used for the synthesis. The model that reproduces the observed spectrum is deemed to be a good representation of the atmospheric conditions of the region from which the observed spectrum was emitted. This is the basic principle behind spectropolarimetric inversion techniques \citep[e.g.][]{SIR, SPINOR, inversion} that are routinely used to obtain an atmospheric model from the observed Stokes profiles. The retrieval of atmospheric parameters using inversions depends on the following: the height range over which a line is sensitive to the atmospheric parameters; the approximations used for the radiative transfer computations \citep[e.g.][]{delaCruzRodriguez17,Smitha21}, how well one can correct for the contamination of Stokes profiles due to the cross talk between them; stray light or smearing by the telescope point spread function \citep[PSF; e.g.][]{OrozcoSuarez07, Michiel12}; and, most importantly, the accuracy of the atomic line parameters used for spectral synthesis.

Together with the local atmospheric conditions, atomic parameters are an important factor in determining the spectral line shape. The oscillator strength $f$ is one of the key parameters that characterizes the spectral line strength. The parameter $f$ is a quantum correction to a classically derived transition probability between two energy levels in an atom, ion, or molecule \citep{Mihalas78}. This parameter is usually given in combination with the statistical weight, $g$, of the lower transition level, as \loggf{} in atomic line databases. The unit measure for \loggf{} is dex (decimal exponent).

Poorly determined atomic parameters remain a problem in spectroscopy, with the most important atomic parameters for studies carried out in local thermodynamic equilibrium (LTE) being, the excitation potential of the lower energy level of the transition, the central wavelength of the spectral line, \loggf{}. The \loggf{} value for a single spectral line can vary greatly between different atomic line databases like NIST \citep[][]{NIST}, Kurucz \citep[][]{Kurucz95} or VALD \citep{VALD}. The origin of these differences mainly lies in the methods used for their determination: laboratory measurements using a spectroscopic furnace \citep[e.g,][]{Blackwell72} or laser-induced fluorescence \citep[e.g,][]{DenHartog05}, theoretical computations based on an atomic model \citep[e.g,][]{Pradhan77}, or using observed solar spectra \citep[e.g,][]{GurtovenkoKostik81, GurtovenkoKostik82, Thevenin89, Thevenin90, Borrero03, Bigot06, Shchukina13, TrellesArjona21}.

Laboratory measurements provide precise \loggf{} values only for isolated and strong lines. Theoretical computations depend exclusively on the distribution of energy levels in the atomic model and therefore have no restrictions on line strength. However, already a small error in the calculation of the energy levels in the atomic model leads to a change in parameters like the central wavelength of the line, collisional rates and consequently to a large uncertainty of the \loggf{} values \citep[][]{Pradhan77}. The advantages and pitfalls of every method are summarised in detail in \cite{Borrero03} and \cite{Shchukina13}. The present work focuses on the inference of atomic parameters based on spatially resolved solar spectra.

An observed solar spectrum permits the inference of atomic parameters of lines whose profiles are unattainable in laboratory setups due to the difficulty in reproducing the physical conditions prevailing in the solar atmosphere. Thus, observed solar spectra provide a unique way of inferring the properties of spectral lines. There are two inversion based methods commonly used in determining the \loggf{} from the solar spectrum. One is the spatially-averaged method, and the other is the pixel-by-pixel method. In the first method, the \loggf{} inference involves modelling the observed spatially averaged solar spectrum with a synthetic spectrum computed using a mean solar atmospheric model by iterative adjustment of \loggf{} values \citep[][and references therein]{GurtovenkoKostik81, GurtovenkoKostik82, Thevenin89, Thevenin90, Borrero03, Shchukina13}. The reliability of the inferred \loggf{}values using this method depends on the quality of the used atmospheric model, adopted abundances of atomic elements, and proper treatment of line broadening to model the observed line profiles accurately. For example, \cite{Bigot06} used the temporally and spatially averaged spectra from a 3D hydrodynamic model of the solar atmosphere to derive \loggf{} from the observed spatially averaged solar spectrum, thus removing the problem of choosing the correct atmospheric model and macro- and micro-turbulent broadening.

\cite{Borrero03} showed that uncertainties in the central wavelength of a line (\dlam{}) must be accounted for in the spatially-averaged method for a reliable retrieval of \loggf{}. They estimated that the uncertainty in the central wavelength from the spatially-averaged method is 5\,m\AA{} at 1\,$\mathrm{\mu m}$ which is comparable to the uncertainty achieved in laboratory measurements of Fe\,I lines \citep{Nave94}. Precise central wavelengths of spectral lines are of significant importance in revising the energy levels of atoms and ions \citep[][]{Borrero03}, identifying new energy levels in atoms and ions \citep{Peterson22}, and for accurate identification of spectral lines in line-rich spectra, especially of iron-group elements \citep{Nave17}.

The pixel-by-pixel method is applied to spatially resolved spectra. In this method, the atomic parameters are inferred independently for each pixel within the observed field-of-view (for further description of this method, see Appendix~\ref{sec:pxl-by-pxl}). \cite{TrellesArjona21} used the pixel-by-pixel method to simultaneously infer the atmospheric parameters along with \loggf{} values for 15 spectral lines around 1.56 $\mathrm{\mu m}$. The final \loggf{} value is the average of all \loggf{} values retrieved from pixels with the best match of the synthetic spectrum to the observed spectrum. The strong cross-talk between atmospheric and atomic parameters can result in a very diverse pool of \loggf{} values, thereby introducing uncertainties in both the inferred atomic as well as atmospheric quantities. 

In the present work, we develop a new inversion method for a reliable determination of both atmospheric as well as atomic parameters for any number of spectral lines simultaneously. During inversions, the atomic parameters for a given spectral line are not allowed to vary at each pixel independently, like in the pixel-by-pixel method, but are allowed to vary only simultaneously in all the pixels. At the same time, this new inversion method allows for the retrieval of the atmospheric parameters in each pixel independently. This new method is part of the new inversion code named \globin{}, and we will refer to this new method in the further text as the \Globin{} method.

The comparison of the spatially averaged observed spectrum and the average synthetic spectrum from a 3D atmosphere shows differences in spectral line strengths in NUV resulting mainly from a poor knowledge of atomic parameters \citep[][]{Tino19}. To retrieve reliable atmospheric parameters from NUV spectra, we will need to first accurately determine the atomic parameters of the spectral lines. The high spectral line density (especially in the NUV) causes the absorption profiles to overlap. This so-called blending of spectral lines is an additional complication for the reliable retrieval of both, the often poorly known atomic parameters of these lines and the atmospheric conditions. We expect that the \Globin{} method will be able to provide a reliable inference of inversion parameters despite the heavy blending between spectral lines.

This work is motivated by the future high spatial and spectral resolution observations in NUV from the instrument \textsc{Sunrise} UV Spectropolarimeter and Imager \citep[henceforth SUSI,][]{SUSI} onboard the next flight of the balloon-borne \textsc{Sunrise} observatory which has successfully flown twice before \citep{Sunrise, Solanki17} and is expected to fly again in June 2024. SUSI will continuously observe the Sun in the near-ultraviolet (NUV) spectral range from $309$\,nm to $417$\,nm. Radiation at these wavelengths is strongly absorbed by the Earth's atmosphere, which makes it hard to observe them from the ground.

The application of the \Globin{} method is not restricted to the spectral region observed by SUSI. This method is general enough to be applied to spectral lines in other wavelength regions as well. Additionally, this method is not limited to the inference of only \loggf{} and \dlam{}. It can easily be extended to infer, e.g., also elemental abundances. However, such an extension is outside the scope of this work and will be addressed in future studies.

Efforts have been made to compile a list of spectral lines that have reliable atomic parameters \citep[e.g.][]{Bigot06,Heiter21} to determine the chemical composition of stellar atmospheres observed by the \textsc{GAIA} mission \citep{GAIA}. Similar to the \Globin{} method, which uses spectra from different pixels to constrain the atomic parameters, spectra from different stars have been used instead to infer the atomic parameters of lines, predominantly in the visible and infrared wavelengths \citep[e.g.][]{Boeche16,ALICCE,LaverickPhD}. We expect that the \Globin{} method would aid in providing reliable atomic parameters of spectral lines that would also be useful to the wider stellar physics community.

Section \ref{sec:global_method_short} introduces the \Globin{} method, and in section\,\ref{sec:test_global}, we demonstrate the potential of this method for inferring \loggf{} and to correct for any errors in the available central wavelength of the spectral lines. The results from the \Globin{} method are compared to the results from the pixel-by-pixel method. For this comparison, we use synthetic multi-line spectra computed from a realistic 3D magnetohydrodynamic (MHD) simulation of the solar atmosphere made with the MPS/University of Chicago Radiative MHD code (MURaM) \citep{Rempel12}. In section\,\ref{sec:standard_method}, we apply the spatially-averaged method to a spatially averaged synthetic spectrum from a quiet-sun MURaM atmospheric model. Here, we test the applicability of the spatially-averaged method for the retrieval of atomic parameters from blended spectral lines. Conclusions are given in section\,\ref{sec:conclusions}.

\section{The \Globin{} method}
\label{sec:global_method_short}

Spectropolarimetric inversion is a degenerate problem where different sets of atmospheric parameters can result in very similar spectra \citep[for a review see][]{inversion}. This implies that a certain degree of cross-talk exists between the atmospheric parameters even when the atomic parameters are well known. When the atomic parameters are also inverted, then this cross talk becomes even larger, especially between those quantities that affect a given spectral line in a similar way, for example, between temperature and \loggf{}, and between line of sight velocity and uncertainties in the central wavelength of the line. 

To circumvent this problem, we have developed the \Globin{} method, where the inference of atmospheric parameters is done individually at every pixel while fitting for only one set of atomic parameters for all pixels in a coupled way. The \Globin{} method uses the Levenberg-Marquardt (LM) optimization algorithm for the inference of the atmospheric and atomic parameters. In the LM optimization scheme, we have to solve the equation:
\begin{equation}
    \mathcal{H} \Delta \mathbf{p_j} = \mathcal{J}^\mathrm{T}\cdot\mathbf{\Delta},
    \label{eq:main_LM}
\end{equation}
where $\mathbf{\Delta}_i = \sqrt{\frac{2}{N-n}}\cdot\frac{w_i}{\sigma_i}\cdot\left( \mathbf{O}_i - \mathbf{S}_i(\mathbf{p_j}) \right)$\footnote{For a comprehensive derivation of the \eq{eq:main_LM}, see \sref{sec:pxl-by-pxl}.}. 

The solution of \eq{eq:main_LM} gives the correction $\Delta\mathbf{p_j}$ to the current set of parameters $\mathbf{p_j}$. The correction to the atomic parameters is determined from differences between observed $\mathbf{O}$ and computed spectra $\mathbf{S(\mathbf{p_j})}$ in all pixels, while the corrections to the atmospheric parameters of individual pixels are determined from the differences between spectra in that pixel alone. This kind of correction is achieved by separating the atmospheric and atomic parameters in the Jacobian matrix of the system $\mathcal{J}$ as:
\begin{equation}
    \mathcal{J} = \left(\  \mathrm{diag}(\mathcal{J}_1^\mathrm{atm}, \dots, \mathcal{J}_N^\mathrm{atm})\ \vert\ \mathcal{J}^\mathrm{atom} \ \right)
\end{equation}
where $\mathcal{J}_i^\mathrm{atm}$ is the Jacobian matrix of the atmospheric parameters in the pixel $i$ $\left(i=\overline{1,N}\right)$ from the observed field of view, and the $\mathcal{J}^\mathrm{atom}$ is the Jacobian matrix of the atomic parameters (spanning all rows in $\mathcal{J}$). The columns of these Jacobian matrices correspond to the response functions of the atmospheric and atomic parameters that describe how the spectrum is altered by changing the inversion parameters. The Hessian matrix $\mathcal{H}$ is computed from the Jacobian matrix as $\mathcal{H}=\mathcal{J}^\mathrm{T}\mathcal{J} + \lambda_\mathrm{M}\cdot \mathrm{diag}(\mathcal{J}^\mathrm{T}\mathcal{J})$, where $\lambda_\mathrm{M}$ is the Marquardt parameter that controls the sampling of the parameters space.

By coupling the atomic parameters for all pixels, the total number of (free) inversion parameters is reduced compared to the pixel-by-pixel method, in which they vary independently for each pixel. The coupling of atomic parameters constrains the sampling of the parameter space by the optimization method and aids in obtaining the correct set of inversion parameters. A more detailed description of the \Globin{} method is given in Appendix \ref{sec:global}.

The idea of coupling the spectra in different pixels originates from the 2D inversion scheme that was proposed in \cite{Michiel12}. There, the author used the telescope PSF to couple the spectra of nearby pixels in order to retrieve inversion parameters. This improved the spatial resolution of the inferred atmosphere while also reducing the impact of noise  on the inferred parameter values \citep[see e.g.][]{Michiel13, Tiwari13, Tiwari15, CastellanosDuran2023}. Additionally, \cite{Jaime19} introduced a spatial regularization method limiting the spatial variation of desired inversion parameters. Both of these techniques were concerned mainly with retrieving atmospheric parameters, whereas in the present work, we focus on atomic line parameters. We have chosen to follow the approach of \cite{Michiel12} and extend the 2D-coupling scheme to fit any type of global, field-of-view independent parameters, in this paper, the atomic parameters \loggf{} and \dlam{}.

This new inversion method is implemented in the \globin{} inversion code written in Python that uses the RH code \citep{RH} for the spectral synthesis.\footnote{The code was adapted to be imported in a Python environment as a pre-compiled module. All the functionalities of the original code are preserved.} RH solves the polarized radiative transfer equation for non-LTE (NLTE) line formation using the multilevel accelerated lambda iteration scheme \citep[][]{RH91}. We opted for the RH code because of its versatility in treating NLTE line formation, partial frequency redistribution and continuum opacity fudge correction \citep[][]{Bruls92}. All these features will be necessary for future analyses, e.g. of NUV spectra from the SUSI instrument. A detailed description of the implementation and functionalities of \globin{} are given in Appendix \ref{sec:globin}.

\section{Comparison between the pixel-by-pixel method and the \Globin{} method}
\label{sec:test_global}

\begin{figure*}
    \centering
    \includegraphics[width=\textwidth]{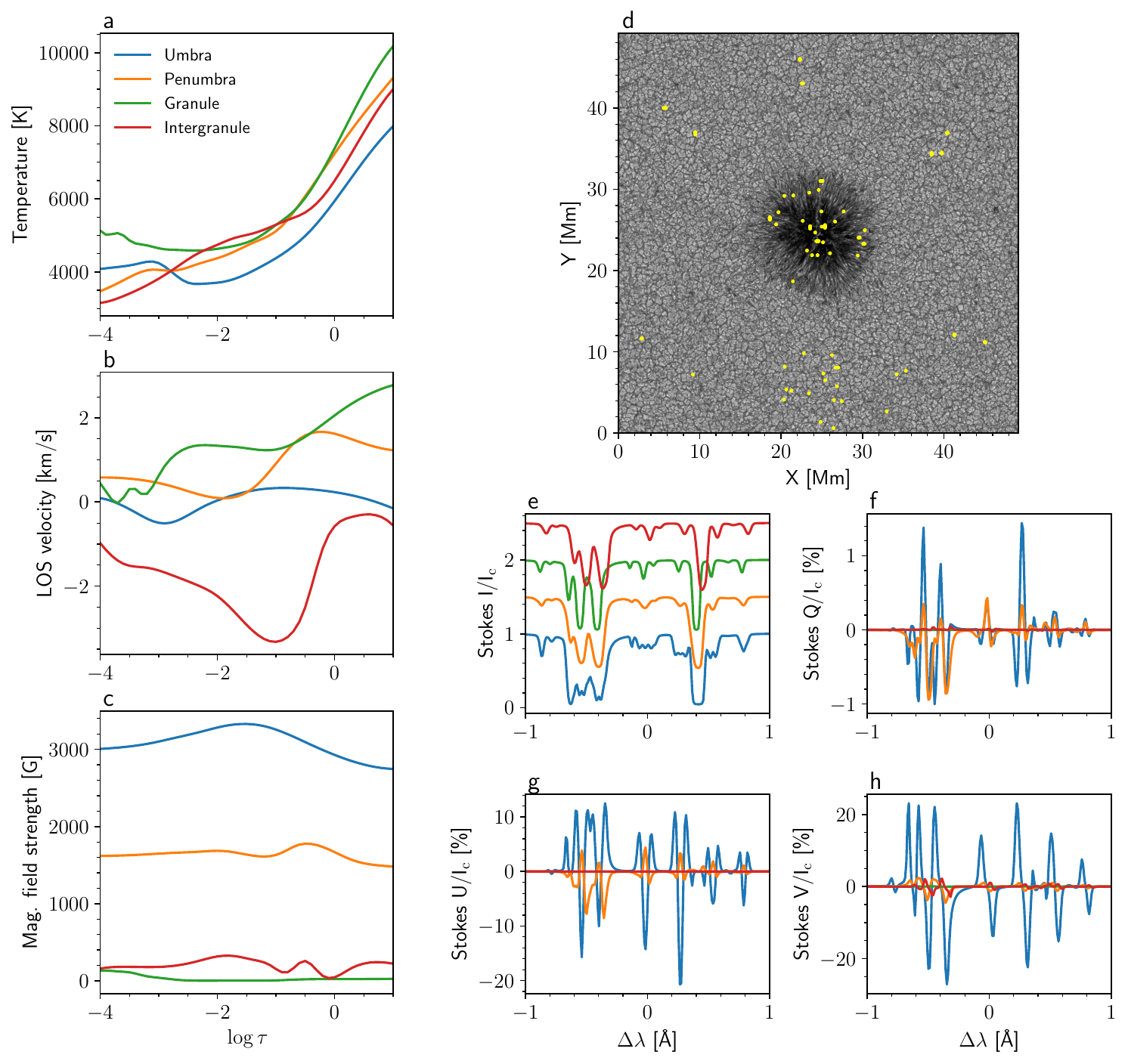}
    \caption{\textit{Panels a-c:} Stratifications of the temperature (a), LOS velocity (b), and magnetic field strength (c) from one pixel for each atmospheric feature taken from the sample of extracted atmospheres. \textit{Panel d}: the continuum intensity from the MHD cube at 4015\,\AA{} with extracted atmospheres annotated with yellow dots. \textit{Panels e-h}: Stokes parameters in the 4015--4017\,\AA{} spectral range from the atmospheres displayed in panels a-c. Wavelengths are given with respect to $4016$\,\AA{}. For easier comparison, the Stokes profiles are normalized to the local continuum intensity $I_c$ and the Stokes $I/I_c$ profiles from different features are shifted vertically.}
    \label{fig:atmospheres}
\end{figure*}

To test the capabilities of the \Globin{} method in comparison to the pixel-by-pixel method, we analyse synthetic Stokes profiles sampling different features such as an umbra, penumbra, granule and intergranular lane, obtained from a 3D MHD snapshot of a sunspot \citep[][see the continuum image in \fig{fig:atmospheres}d]{Rempel12} simulated using the MURaM code \citep{MURAM}. This selection guarantees a diverse sample of temperature, line-of-sight (LOS) velocity and magnetic field strength stratifications (\fig{fig:atmospheres}a-c) resulting in very different Stokes profiles (\fig{fig:atmospheres}e-h). This diversity leads to line profiles, especially of weaker lines, that are prominent in some features, but not visible in others. In particular, the response of lines to a change in the atomic parameters is different for the different atmospheres. This result is beneficial for reducing the cross-talk between atomic and atmospheric parameters in the \Globin{} method.

We will test the applicability of the pixel-by-pixel method and the \Globin{} method to determine the atomic parameters \loggf{} and to correct for possible errors in the central wavelength (\dlam{}) of spectral lines in the 4015--4017\,\AA{} range, containing several blended lines. This spectral region was chosen because it is covered by the Hamburg atlas spectrum \citep{Neckel84}, it has many blended spectral lines, and it lies within the wavelength range covered by the SUSI instrument. The same spectral range will also be considered later while testing the spatially-averaged method (section \ref{sec:standard_method}).

Spectral line information for the considered spectral region are taken from the Kurucz line list \citep[][]{Kurucz95}. There are many lines in the Kurucz line list that are fairly weak and are expected to have an insignificant effect on the spectrum. Including all of them will only extend the computing time. We consider only those spectral lines with a line core depth of at least 1\% of the local continuum intensity. We found 18 such lines in this region. Their parameters are given in \tab{tab:lines}. The blending factor for each line is estimated using the method in \cite{Laverick17}. This factor quantifies the extent of overlap between the core of a given absorption line with its neighbouring lines. The smaller the blending factor is, the smaller the amount of overlap. 

The Stokes profiles are computed at the disk centre ($\mu=1$) under the LTE approximation using \globin{}. The computed spectra are normalized using the continuum intensity computed from the HSRA atmospheric model\footnote{Any other mean atmospheric model, such as FAL-C or HOLMUL, could be used to compute the continuum intensity.} \citep[][]{HSRA}. The differences between the \Globin{} method and the pixel-by-pixel method are tested under ideal conditions by disregarding observational effects such as stray light, the finite resolving power of spectrographs and noise.

In the inversion, we fitted for the height stratification of the atmospheric parameters: temperature, LOS velocity, and magnetic field vector. The temperature is inferred using four nodes placed at optical depth values of \logtau{}$=(-2.5, -1.5, -0.5, 0.4)$ while the remaining atmospheric parameters are inferred at three nodes, \logtau{}$=(-2.2, -1.1, 0)$. Here and in the rest of the paper, the optical depth scale is computed at the reference wavelength of 5000\,\AA. The initial values for the atmospheric parameters are uniform across all pixels. 

Among the atomic parameters, we considered \loggf{} and the central wavelength shift \dlam{} of all lines as free parameters, except for line 13, whose parameters were kept fixed during inversions in both the pixel-to-pixel and \Globin{} methods. Line 13 is the strongest and the least blended within the spectral range. Fixing the atomic parameters of one spectral line is necessary for the absolute wavelength and LOS velocity estimate, and additionally improves the inference reliability of atomic parameters for other lines in the spectral region. 

The initial values for the atomic parameters are randomized with a Gaussian distribution centred at the exact value (i.e., the value used for computing the synthetic spectra from the simulations) with a standard deviation of 0.2\,dex for \loggf{} and 5\,m\AA{} for the \dlam{} parameter. This initialization is compatible with the expected uncertainties in these parameters and reflects the realistic situation we have to deal with when applying the method to observed spectra. We limit the allowed ranges for \loggf{} and \dlam{} to $[-2.0,1.5]$\,dex and $[-30,22]$ m\AA{} with respect to the exact values. These limits are in the range of the expected uncertainties in the atomic databases and can be adjusted for each spectral line independently. The asymmetric boundaries are chosen to remove any possibility for a zero average value, which could occur from averaging inferred values from symmetric boundaries.

To compare results from the \Globin{} method with the pixel-by-pixel method, we performed the inversions in three different modes:

\begin{itemize}
    \setlength\itemsep{1em}
    \item[] \textbf{Mode\,1:} Only atmospheric parameters are inverted for each pixel individually. The atomic parameters are fixed to the values used to compute the reference spectra. This inversion allows us to retrieve the best possible stratification of atmospheric parameters with the given node settings and initial parameter values since all atomic parameters are assumed to be accurately known.
    
    \item[] \textbf{Mode\,2:} Both atomic and atmospheric parameters are inverted for each pixel independently (pixel-by-pixel method). Here, the inversion retrieves different atomic parameters for every spectral line and every pixel.
    
    \item[] \textbf{Mode\,3:} Atmospheric parameters vary between pixels, while atomic parameters are inverted globally (i.e., the \Globin{} method). Only one set of atomic parameters (\loggf{} and \dlam{}) is obtained per spectral line.
\end{itemize}

Mode\,1 inversion results are used as a reference to which the inversion results from mode\,2 and 3 are compared. Any large deviations in the retrieved atmospheric parameters in mode\,2 and 3 in comparison to the retrieved atmospheric parameters in mode\,1 can likely be attributed to the poor atomic parameters in these two modes. All three modes were run with the same initial values of the atmospheric parameters, and the same initial atomic parameters in mode\,2 and 3.

\begin{figure}
    \centering
    \includegraphics[width=\linewidth]{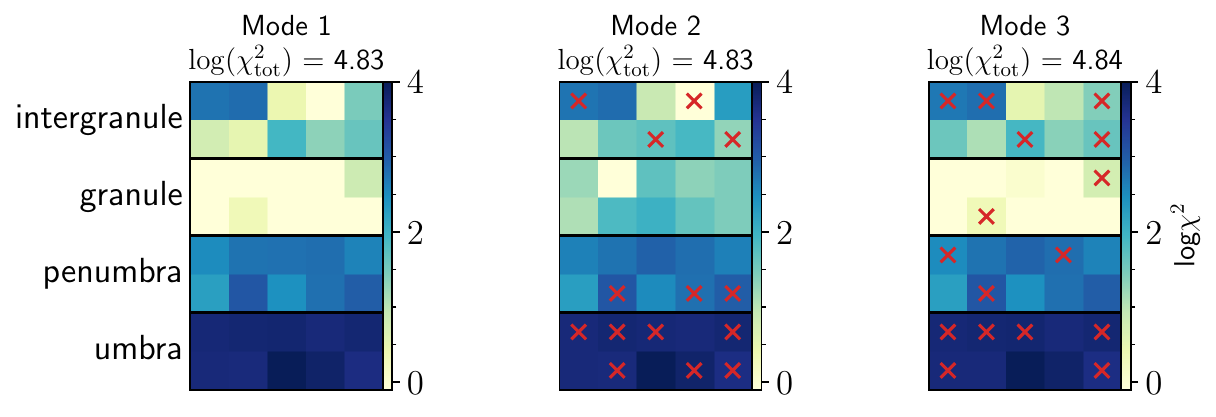}
    \caption{$\chi^2$ values for all three modes on a logarithmic scale. The total $\chi^2$ value, the sum of the $\chi^2$ values from all pixels, is displayed at the top of each panel. Pixels marked with a red cross in panels of mode\,2 and 3 correspond to pixels with a lower $\chi^2$ than the $\chi^2$ value in mode\,1.}
    \label{fig:chi2}
\end{figure}

The $\chi^2$ values\footnote{For direct comparison, $\chi^2$ is calculated for each pixel as $\chi^2=\frac{1}{N}\sum_{i=1}^N w_i^2\left(\mathbf{O}_i - \mathbf{S}_i\right)^2$ where $N$ is the number of wavelengths (running over each Stokes parameter), $w_i$ is the weighting of each Stokes component and each wavelength, $\mathbf{O}$ is a synthetic Stokes spectrum from selected pixels and $\mathbf{S}$ is an inverted Stokes spectrum. Both $\mathbf{O}$ and $\mathbf{S}$ are normalized to the local Stokes $I$ continuum value. The same weights were used in all three modes.} for all selected atmospheres (referred to as pixels from now on) from all three modes are displayed in \fig{fig:chi2}. Note that these pixels are not spatially connected (yellow dots in \fig{fig:atmospheres}d). Inversion mode\,1 shows a very low $\chi^2$ value for pixels within the granules, whereas the umbral and penumbral pixels show higher values. The convergence to lower $\chi^2$ in these pixels could be improved by changing the initial parameters or changing the convergence parameters of the LM algorithm. However, this does not guarantee that the $\chi^2$ value in other pixels would also improve simultaneously.

The inversion mode\,2 and \,3 show a very good fit to the Stokes spectra, which is comparable to mode\,1, with slight differences in individual pixels. Those pixels in mode\,2 and 3 with $\chi^2$ value lower than in mode\,1 are marked with red crosses in \fig{fig:chi2}. With a larger number of free parameters, the inversion algorithm manages to find a lower minimum in the $\chi^2$ hypersurface. 

The quality of the retrieved atmospheric parameters stratification in each pixel for all three inversion modes is quantified using the root-mean-square-deviation (RMSD) estimator. The parameter's RMSD ($P_\mathrm{RMSD}$) is defined as the root-mean-square difference between the retrieved parameter stratification ($P_\mathrm{inv}$) and the original parameter stratification in the MHD cube ($P_\mathrm{MHD}$) at all depths between the highest and lowest nodes on the interpolated \logtau{} grid: 
\begin{equation}
    P_{{\rm RMSD}} = \sqrt{\frac{1}{N_\mathrm{d}} \sum_{i=1}^{N_\mathrm{d}} \left(P_{\mathrm{inv}_i} - P_{\mathrm{MHD}_i}\right)^2},
\end{equation}
where $N_\mathrm{d}$ is the number of fine grid points in the atmosphere between the lowest and the highest node of the parameter $P$. The extrapolated depth points above the highest and below the deepest node were excluded from the computation of the $P_\mathrm{RMSD}$ to focus on the region of the atmospheres where the lines are formed.

\begin{figure}
    \centering
    \includegraphics[width=\linewidth]{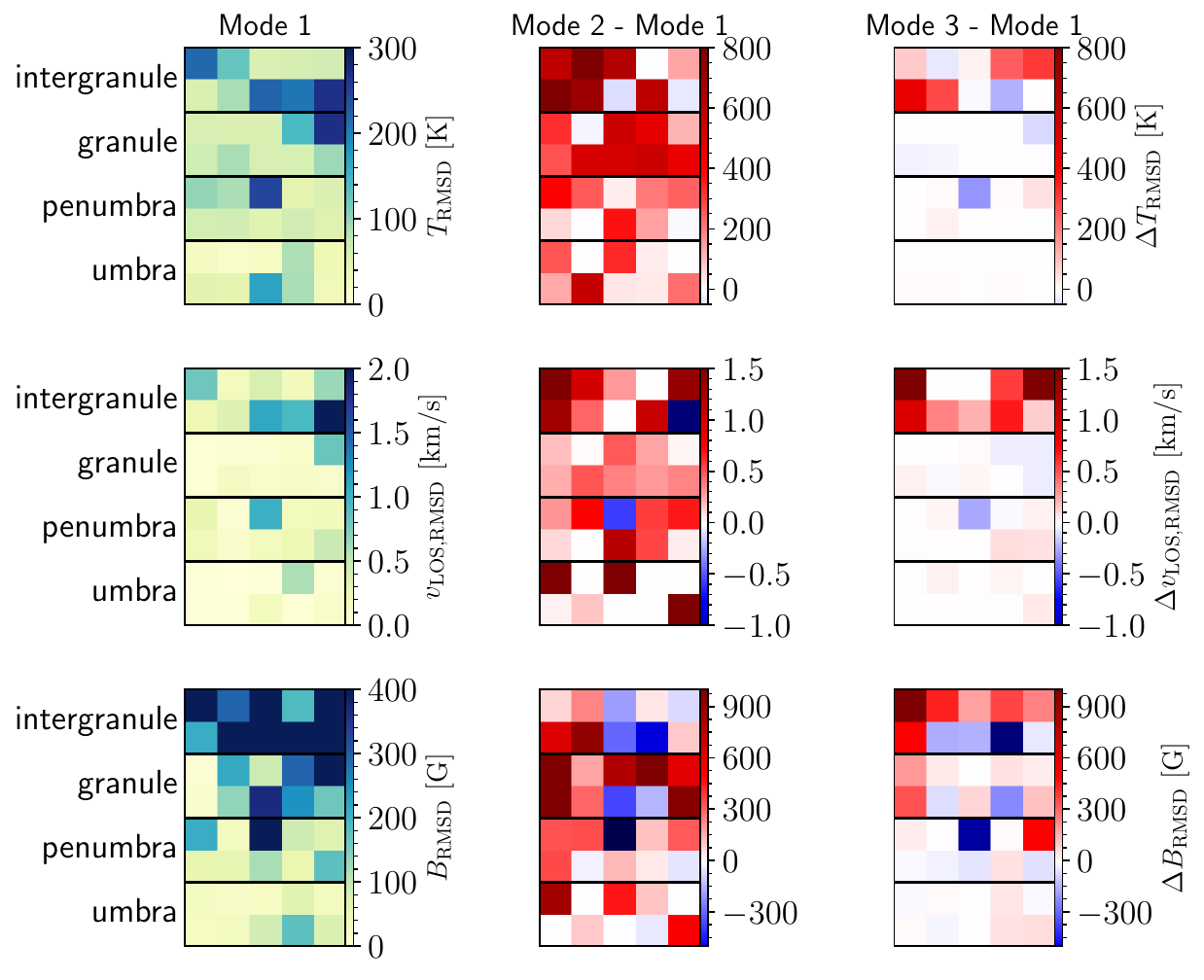}
    \caption{The root-mean-square deviation of the retrieved atmospheric parameters stratification calculated from the highest to the lowest nodes in each inversion mode for each pixel. The first column shows the $P_\mathrm{RMSD}$ for mode\,1, while the other two columns display differences in $P_\mathrm{RMSD}$ from mode\,2 and mode\,3 from mode\,1. The $P_\mathrm{RMSD}$ measures are given for temperature (first row), LOS velocity (second row) and the magnetic field strength (third row). }
    \label{fig:rms}
\end{figure}

The $P_\mathrm{RMSD}$ values for the temperature, LOS velocity and magnetic field strength for each pixel are displayed in \fig{fig:rms}. The first column shows the $P_\mathrm{RMSD}$ for mode\,1 while the last two columns display the differences in $P_\mathrm{RMSD}$ for mode\,2 and 3 relative to mode\,1 ($\Delta P_\mathrm{RMSD}$ value). Negative differences indicate a better retrieval of atmospheric parameters in mode\,2 and 3 compared to the atmospheric parameters retrieved in mode\,1. Overall, mode\,3 shows a better retrieval of atmospheric parameters compared to mode\,2, and the values retrieved are comparable to mode\,1 inversion results.

A temperature offset is observed in some pixels in mode\,2 that are related to umbra and penumbra features (see \fig{fig:rms}). In these pixels, the LM algorithm manages to achieve the same or lower $\chi^2$ in comparison to the one obtained in mode\,1. This is because, at every pixel in mode 2, the code has the freedom to adjust \loggf{} and other atmospheric parameters independently to fit the line profile. When the algorithm chooses to over-tweak the \loggf{} to get the desired fit, this can result in an offset of parameters like temperature, which affects a line profile in a way similar to the \loggf{}. This so-called cross-talk between atomic and atmospheric parameters can result in large deviations in the retrieved values, while achieving a satisfactory fit to the Stokes profiles. One of the reasons for this is the use of a simple four-node representation for the temperature stratification in the inversions. It is possible that such a simple model is not adequate to accurately represent the complex height stratification found in the MHD atmospheres. On the other hand, the retrieved atmospheric parameters from mode\,3 show a smaller deviation from the values retrieved in mode\,1, which indicates that the coupling between pixels weakens the cross-talk between the temperature and the \loggf{}.

\begin{figure}
    \centering
    \includegraphics[width=\linewidth]{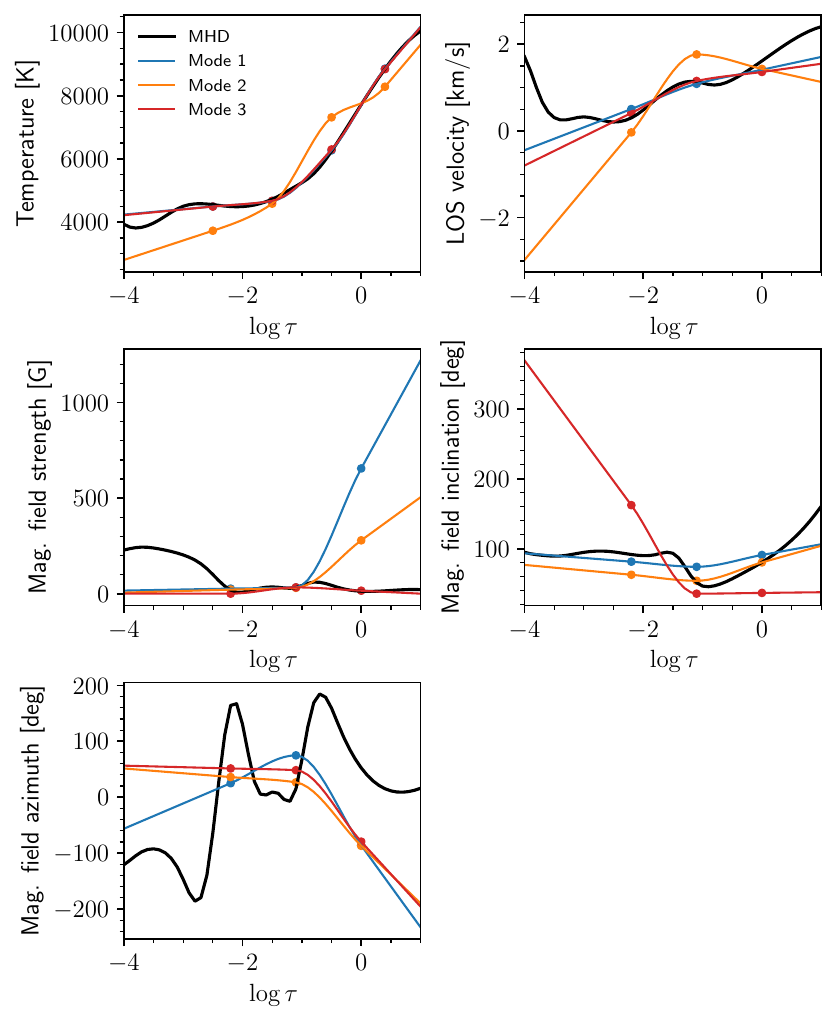}
    \caption{Stratifications of various atmospheric parameters for a granule atmosphere. Curves with different colours represent the stratification of the original MHD simulation and different inversion modes (see legend in the top-left panel). Circles represent the node positions.}
    \label{fig:atmos_compare}
\end{figure}

\begin{figure}
    \centering
    \includegraphics[width=\linewidth]{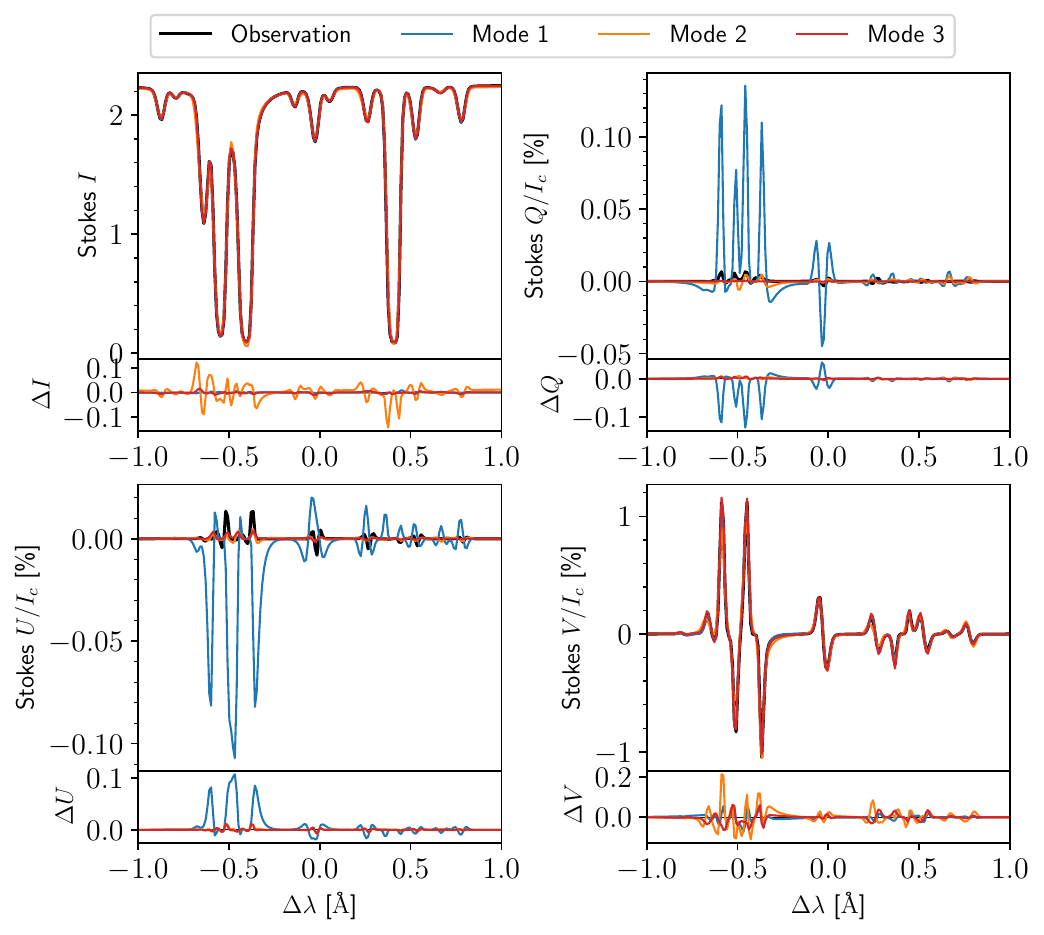}
    \caption{The comparison of Stokes spectra in all inversion modes for a granule atmosphere whose atmospheric stratification is displayed in \fig{fig:atmos_compare}.}
    \label{fig:spec_compare}
\end{figure}

A comparison of the retrieved parameters stratification in all three modes to the MHD stratification for a granule atmosphere is displayed in \fig{fig:atmos_compare}. This figure exemplifies the highly complex stratifications typically found in MHD simulations. The corresponding synthetic Stokes spectra and the best fit for the three modes are displayed in \fig{fig:spec_compare} where mode 3 shows the best fit to the observed spectrum.

\begin{figure*}
    \centering
    \includegraphics[width=\textwidth]{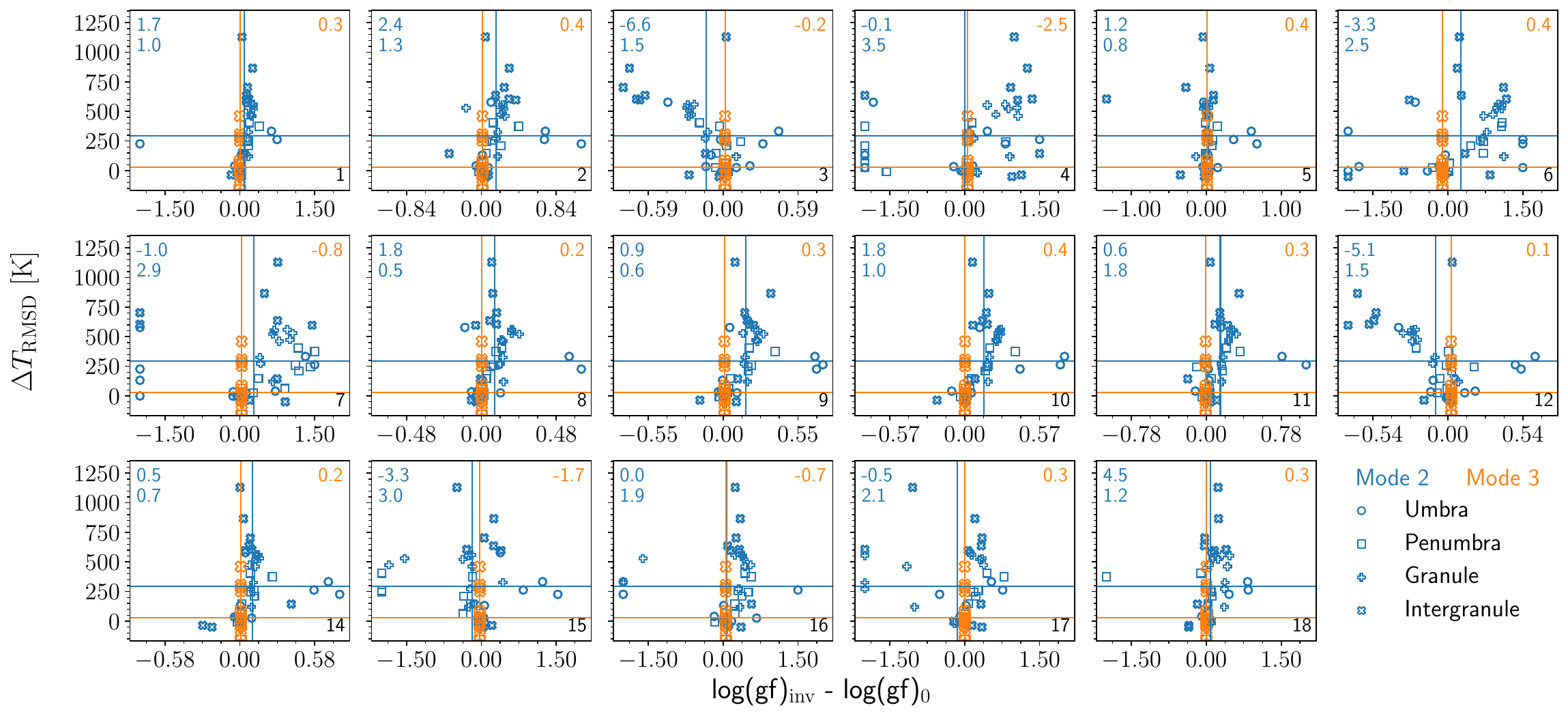}
    \caption{Scatter plots of the \loggf{inv}-\loggf{0} ($x$-axis) versus $\Delta T_\mathrm{RMSD}$ values ($y$-axis) for mode\,2 (in blue) and mode\,3 (in orange). Each subplot corresponds to one spectral line with the line number from \tab{tab:lines} given in the lower-right corner. The values from each pixel are represented with different symbols: circle (umbra), square (penumbra), plus (granule) and cross (intergranule). Horizontal and vertical lines mark the average of $\Delta T_\mathrm{RMSD}$ and \loggf{inv}-\loggf{0} in each inversion mode, respectively. The values in the upper left corner of each panel are the mean and standard error of \dlam{} in mode\,2, and in the upper right corner is the inferred \dlam{} in mode\,3 in units of m\AA{}. The standard error of \dlam{} is zero in mode\,3 since it retrieves a unique global value.}
    \label{fig:rms_vs_loggf}
\end{figure*}

\begin{figure}[h!]
    \centering
    \includegraphics[width=\linewidth]{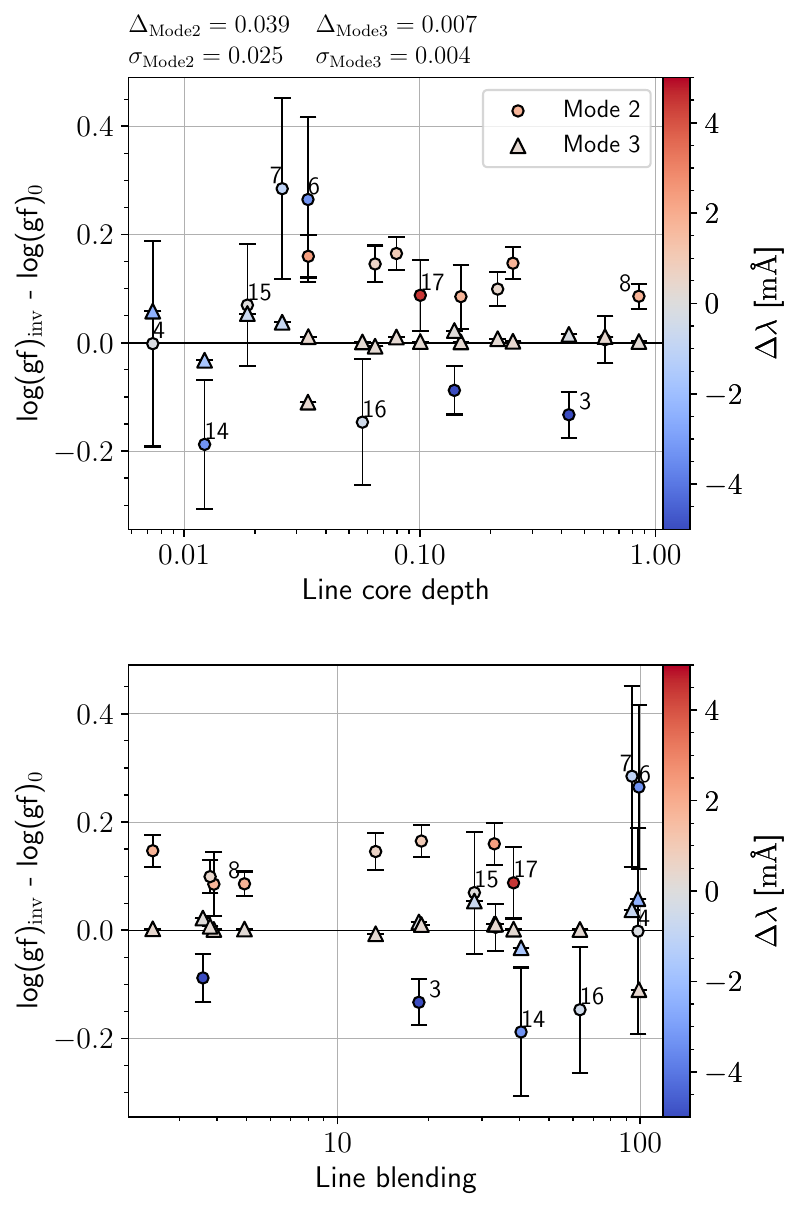}
    \caption{Comparison of the atomic parameters \loggf{} and \dlam{} for mode\,2 and 3 inversions. \textit{Top panel:} The $y$-axis shows the difference between inferred \loggf{inv} and exact \loggf{0} values where on the $x$-axis we plotted the line core depth. The averaged values in mode\,2 are marked in circles and the values from mode\,3 in triangles. Markers are colored based on the \dlam{} parameter. The top of the panel lists mean $\Delta$ and standard error $\sigma$ of \loggf{inv} - \loggf{0} for each mode. The error bars are displayed only for mode\,2, and they mark the 1$\sigma$ level. Mode\,3 retrieves a unique global value for each line, and we do not have any measure of error in this case. \textit{Bottom panel:} The comparison of \loggf{inv}-\loggf{0} with the line blending factor listed in \tab{tab:lines}. Markers shape and color have the same meaning as in the top panel.}
    \label{fig:loggf_diff}
\end{figure}

The comparison between the retrieved \loggf{inv} in mode\,2 and 3 and the \loggf{0} is shown in \fig{fig:loggf_diff}. The mode\,2 results represent the mean value from the \loggf{} values of all considered pixels. 
This not only improves the statistical significance for the mode\,2 results but also allows for a fair comparison with the mode\,3 results, which always takes into account all considered pixels. A preliminary analysis shows that no significant improvement in the results from mode\,2 can be achieved by increasing the number of pixels. Also, it is to be expected that the optimal number of pixels required for each of these modes will depend on the number of lines for which the atomic parameters are to be determined and the amount of line blending. The influence of the chosen number of pixels on the results from both, mode\,2 and 3, will be investigated in detail in a follow-up study.

As a measure of the quality of the \loggf{} inference for mode\,2, we use the mean and the standard error of the \loggf{inv} - \loggf{0}. The mode\,2 inversion manages to retrieve the \loggf{} values to an accuracy of $0.025$\,dex of the exact value, while mode\,3 performs better, with an accuracy of only $0.004$\,dex. These values were computed by averaging the retrieved values weighted by their line core depth, giving larger weight to stronger and less blended lines. The standard error of \dlam{} is below 1\,m\AA{} for both modes.

Spectral line 4 has the largest standard error of \loggf{} in mode\,2 and the mean \loggf{} close to the expected value (\fig{fig:loggf_diff}, see also \fig{fig:rms_vs_loggf}). This is likely a result of the severe blending with line 3: changes in \loggf{} of line 4 will not produce a significant signature in the spectrum, which hinders the inversion in retrieving a reliable value. A similar situation also holds for line 7, which forms a blend with line 8. \fig{fig:rms_vs_loggf} shows that the uncertainty of the central wavelength of the spectral line can have a large impact on the retrieval of the exact \loggf{} value. This is especially seen in line 4. 

The problem in retrieving atomic parameters caused by the blending of spectral lines is illustrated by comparing the results for lines 15 and 16. \fig{fig:loggf_diff} shows that line 15 is weaker while line 16 is stronger than expected in mode\,2. However, line 15 also shifts to shorter wavelengths, moving it further away from line 16. To reproduce the synthetic profile, the inversion code decreases the \loggf{} of line 15 and increases it for line 16. This interplay between the \loggf{} values of these lines is visible in \fig{fig:rms_vs_loggf}. This problem is alleviated in mode\,3, where no such large difference in retrieved \loggf{} or \dlam{} is observed. A similar explanation for the strongly blended pair of lines 17 and 18 also holds. This result clearly shows the benefits of using the \Globin{} method to retrieve reliable atomic parameters for blended lines compared to the pixel-by-pixel method.

\section{Test of the spatially-averaged method}
\label{sec:standard_method}

The standard method of inferring atomic parameters from observations relies on synthesizing a spectrum from a representative  atmospheric model and adjusting only the atomic parameters until the best fit to a spatially averaged solar or stellar spectrum is achieved. We call this method the `spatially-averaged method'. The information content in a single spectrum is lower than when using many spectra emerging from different atmospheric conditions. Therefore, the method works best when applied to strong and isolated spectral lines such as those found in the visible and the infrared spectral regions \citep[e.g.,][]{GurtovenkoKostik81, GurtovenkoKostik82, Thevenin89, Thevenin90, Borrero03, Bigot06}. When the lines are blended (i.e., their absorption profiles overlap in wavelength), the method requires simultaneous treatment of all blended spectral lines \citep{Borrero03} or a proper de-blending of the spectral line of interest \citep{Shchukina13}.

We perform a test of the spatially-averaged method for the same spectral region as was used in section~\ref{sec:test_global}, i.e., 4015--4017 \AA{}, containing 18 spectral lines (\tab{tab:lines}). Here, we computed a spatially averaged spectrum from another quiet-sun MURaM simulation having a mean magnetic field of strength of $50$\,G. We expect that this spatially averaged spectrum is rather similar to the observed  average spectrum of the quiet Sun.

\begin{table*}
    \centering
    \caption{Atomic line parameters from 4015--4017\,\AA{} range.}
    \begin{tabular}{c l c c c c r c c c c c c c}
    \toprule[1.5pt]
    No & element & $\lambda_0$ [\AA{}] & $E_i$ [eV] & $\mathsf{g}_\mathrm{eff}$ & $A$ & $b$ [\%] & $D_\mathrm{core}$ & \loggf{0} & \loggf{mode2} & \loggf{mode3} & \loggf{spat} \\ \midrule[1.5pt]
    1 & Fe\,I   & 4015.139 & 4.1631 & 0.666 & 7.44\tablefootmark{b} & 3.9  & 0.149 & -2.172 & -2.087 & -2.170 & -2.166 \\ 
    2 & Co\,I   & 4015.219 & 2.7849 & 0.484 & 4.92\tablefootmark{a} & 32.9 & 0.033 & -1.806 & -1.646 & -1.795 & -1.610 \\ 
    3 & Ti\,I   & 4015.373 & 2.0850 & 0.515 & 4.99\tablefootmark{a} & 18.5 & 0.428 & -0.084 & -0.217 & -0.069 & -0.105 \\ 
    4 & V\,I    & 4015.398 & 2.0618 & 0.995 & 4.00\tablefootmark{a} & 98.3 & 0.007 & -0.857 & -0.859 & -0.799 & -0.221 \\ 
    5 & Fe\,I   & 4015.465 & 4.0434 & 1.407 & 7.44\tablefootmark{b} & 33.3 & 0.610 & -0.781 & -0.775 & -0.770 & -0.787 \\
    6 & Ni\,II  & 4015.474 & 3.9028 & 0.619 & 6.25\tablefootmark{a} & 99.0 & 0.033 & -2.419 & -2.154 & -2.529 & -2.429 \\
    7 & Nd\,II  & 4015.545 & 1.6766 & 0.886 & 1.50\tablefootmark{c} & 93.9 & 0.026 &  0.080 &  0.365 &  0.118 &  0.589 \\ 
    8 & Fe\,I   & 4015.605 & 4.0434 & 1.394 & 7.44\tablefootmark{b} & 4.9  & 0.849 & -0.515 & -0.429 & -0.513 & -0.542 \\ 
    9 & Ce\,II  & 4015.875 & 1.0083 & 0.881 & 1.55\tablefootmark{c} & 18.9 & 0.079 & -0.087 &  0.078 & -0.076 &  0.018 \\
    10 & Fe\,I  & 4015.986 & 4.1235 & 2.000 & 7.44\tablefootmark{b} & 2.4  & 0.248 & -1.928 & -1.781 & -1.925 & -1.922 \\ 
    11 & Ni\,I  & 4016.068 & 3.9735 & 1.388 & 6.25\tablefootmark{a} & 13.3 & 0.064 & -1.870 & -1.724 & -1.876 & -1.760 \\ 
    12 & Ti\,I  & 4016.274 & 2.0658 & 1.762 & 4.99\tablefootmark{a} & 3.6  & 0.140 & -0.714 & -0.802 & -0.691 & -0.733 \\ 
    13 & Fe\,I  & 4016.419 & 3.1775 & 0.682 & 7.44\tablefootmark{b} & 1.0  & 0.865 & -1.600 &        &        & -1.633 \\ 
    14 & Fe\,I  & 4016.541 & 2.6399 & 1.197 & 7.44\tablefootmark{b} & 3.7  & 0.213 & -3.513 & -3.414 & -3.505 & -3.498 \\ 
    15 & Mn\,I  & 4016.658 & 4.1926 & 0.035 & 5.39\tablefootmark{a} & 40.4 & 0.012 & -1.160 & -1.348 & -1.192 & -0.876 \\ 
    16 & Co\,II & 4016.685 & 3.0177 & 1.293 & 4.92\tablefootmark{a} & 28.3 & 0.018 & -2.905 & -2.835 & -2.851 & -3.197 \\
    17 & Fe\,I  & 4016.792 & 4.1195 & 2.002 & 7.44\tablefootmark{b} & 63.2 & 0.057 & -2.576 & -2.723 & -2.574 & -2.357 \\ 
    18 & Co\,I  & 4016.793 & 3.5124 & 1.213 & 4.92\tablefootmark{a} & 38.2 & 0.100 & -0.547 & -0.459 & -0.545 & -0.792 \\
    \bottomrule[1.5pt]
    \end{tabular}
    \tablefoot{The columns represent line number in the spectral region used for analysis (see \fig{fig:muram_inverted_spec_compare}), chemical element, central wavelength $\lambda_0$, the energy of lower level $E_i$, blending factor $b$, line core depth $D_\mathrm{core}$ effective Land\'e factor $\mathsf{g_\mathrm{eff}}$, element abundance $A$, \loggf{0} parameter and inferred \loggf{} in mode\,2 averaged over all pixels (\loggf{mode2}), mode\,3 (\loggf{mode3}) and using the spatially-average method (\loggf{spat}). All atomic data are from Kurucz's database, except for the Land\'e factor and abundances. The effective Land\'e factors are taken from the VALD database, except for line 10, which is computed assuming LS coupling. \tablefoottext{a}{\citet{GrevesAnders91}}\tablefoottext{b}{\citet{Asplund00}}\tablefoottext{c}{\citet{AndersGrevesse89}}.}
    \label{tab:lines}
\end{table*}

We opted to infer a representative 1D atmospheric model from this average spectrum, assuming that the atomic parameters are known and kept fixed during the inversion. A similar approach is also followed in \cite{Borrero03}. Additionally, the Land\' e factors of all synthesized lines are set to zero to 
remove the effect of the broadening introduced by the magnetic field. This simplifies our inferred atmospheric model by eliminating the need to infer the magnetic field vector.

We use the \globin{} code to infer the temperature $T$ in four nodes at $\log\tau=(-2.2, -1.5, -0.8, 0)$, line-of-sight velocity $\mathrm{v}_\mathrm{LOS}$ in two nodes at $\log\tau=(-2.2,0)$, micro-turbulent velocity $\mathrm{v_{mic}}$ in three nodes at $\log\tau=(-2.2, -1.1, 0)$ and depth-independent macro-turbulent velocity $\mathrm{v}_\mathrm{mac}$. The $\mathrm{v_{mac}}$ and depth-dependent $\mathrm{v_{mic}}$ allow us to properly model the broadening of spectral lines resulting from the spatial averaging of spectra from granules and intergranular lanes with different brightness and surface area coverage, and harbouring oppositely directed flows. The spatial averaging also introduces a height dependence of the line-of-sight velocity, which is considered by fitting $\mathrm{v_{LOS}}$ with two nodes.

\begin{figure*}[t!]
    \centering
    \includegraphics[width=\textwidth]{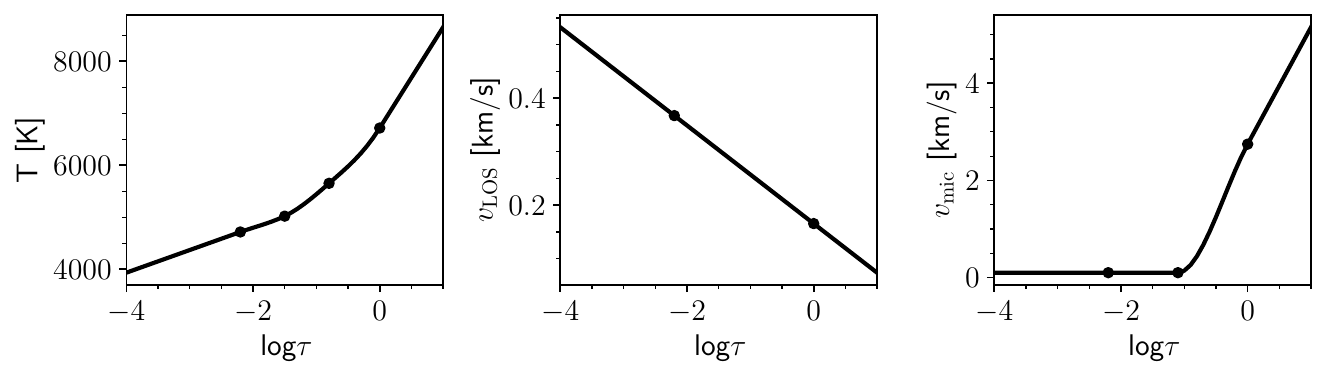}
    \caption{The inferred atmospheric model (temperature, line of sight velocity and micro-turbulent velocity) from the spatially averaged spectrum synthesised from the quiet-sun MURaM cube.}
    \label{fig:muram_inferred_atmos}
\end{figure*}

\begin{figure}
    \centering
    \includegraphics[width=0.45\textwidth]{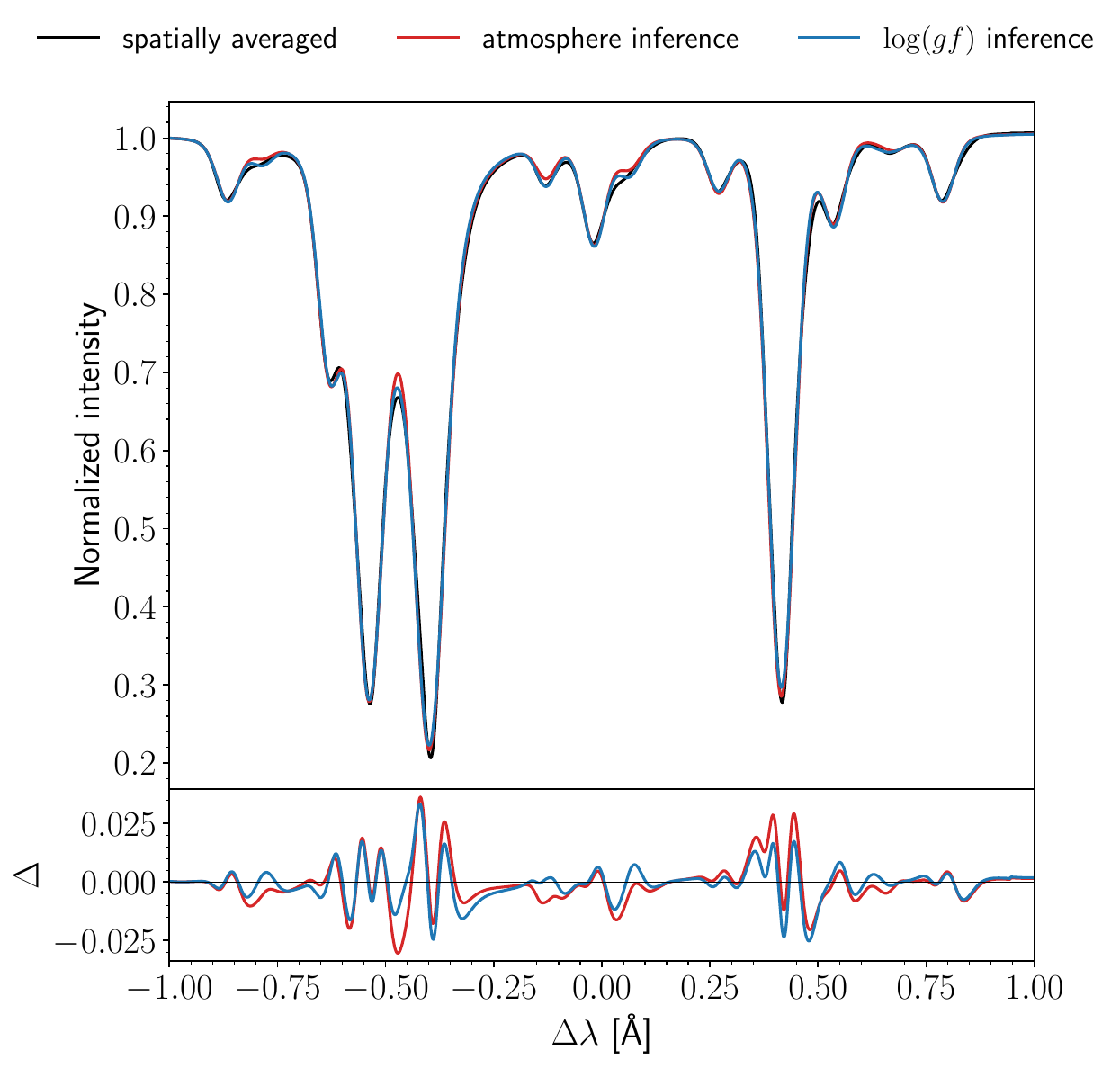}
    \caption{Comparison between spatially averaged spectrum from the quiet-sun MURaM cube (black line), the spectrum from the inference of atmospheric model (red line) and the spectrum from the inference of atomic line parameters (blue line). In the bottom panel, we show the difference between spectra compared to the spatially averaged one.}
    \label{fig:muram_inverted_spec_compare}
\end{figure}

The inferred atmospheric parameters are displayed in \fig{fig:muram_inferred_atmos} with $\mathrm{v_{mac}}=2.05$\,km/s, where the comparison of the spatially averaged and inverted spectra is displayed in \fig{fig:muram_inverted_spec_compare}. The retrieved micro-turbulent velocity is in accordance with what is generally found in any 1D atmospheric model, such as FAL-C \citep{FALC}, with higher values in the deepest layer that decrease with height in the photosphere to values below 1\,km/s. The retrieved line-of-sight velocity shows a small gradient, indicating the need to model the asymmetries in line profiles in the spatially averaged spectrum. 

The comparison of spectra in \fig{fig:muram_inverted_spec_compare} shows that the atmospheric model produces a good match in the cores of many lines, with some difficulties in reproducing the cores of very weak lines. The alteration of atmospheric parameters at heights to which the cores of these lines are sensitive does not significantly improve the $\chi^2$ value. This example shows that it is impossible to simultaneously reproduce the entire profiles of different lines in a spatially averaged spectrum using a 1D atmospheric model. In a more rigorous examination, \cite{Han11} showed that 1D atmospheric models, constructed by averaging the individual atmospheres within a simulated 3D MHD atmosphere of the Sun, do not reproduce the average spectrum emerging from the 3D MHD atmospheric model very well. This result is a consequence of the radiative transfer equation being non-linear, which leads to a non-linear mapping of atmospheric parameters onto the spectrum. Additionally, \cite{Han11} found that if an atmospheric model reproduces the spectrum in a specific wavelength range, it need not reproduce other wavelength ranges equally well.

We have experimented with different numbers of nodes in micro-turbulent velocity and temperature, but none of the models provided as good a match to the spatially averaged spectrum as the chosen one (see Fig.~\ref{fig:muram_inferred_atmos} and \ref{fig:muram_inverted_spec_compare}), which is in agreement with the findings by \cite{Han11}. Therefore, with all the limitations of a 1D atmospheric model, we deemed that the spectrum from the inferred atmospheric model is the best possible representation of the spatially averaged spectrum from the quiet-sun MURaM cube (or at least very close to such a representation). 

We then proceed to infer the atomic line parameters from the spatially averaged spectrum using the inferred atmospheric model. We assume that the \loggf{} parameters of all lines are unknown and need to be retrieved using the \globin{} code. In this test, we assumed the central wavelengths of the lines to be known to ensure that there is no additional cross-talk between any uncertainty in the wavelengths with the \loggf{} parameter. The initial \loggf{} values were randomized around the exact values, as was done in \sref{sec:test_global}. 


\begin{figure}
    \centering
    \includegraphics[width=\linewidth]{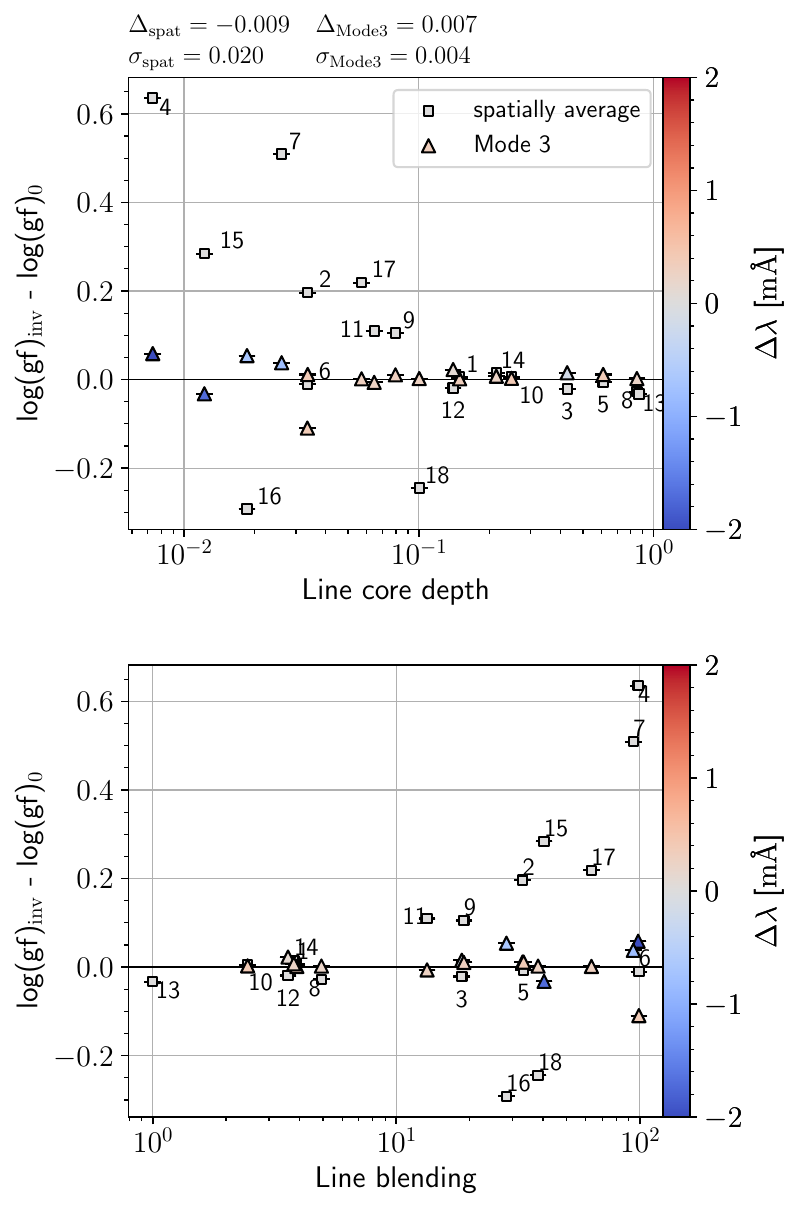}
    \caption{\textit{Top panel:} The difference between inferred and expected \loggf{} values versus the line core depth from the spatially-averaged method (squares) and from mode 3 in section~\ref{sec:test_global} (triangles). The colours of the symbols correspond to the wavelength shift \dlam{} with respect to the central wavelength from \tab{tab:lines} (only for mode 3 results). \textit{Bottom panel:} The difference between inferred and expected \loggf{} values versus the line blending factor from \tab{tab:lines}. Colour coding is the same as in the top panel.}
    \label{fig:muram_atom_pars}
\end{figure}



The inferred values for both the spatially averaged and the \Globin{} method from section~\ref{sec:test_global} are displayed in \fig{fig:muram_atom_pars}. Both methods can reproduce the \loggf{} and $\Delta\lambda$ values reasonably well for strong (line core depth $\ge 0.1$) and unblended (line blending $\leq 10$) spectral lines. But even for the stronger lines, the \Globin{} method produces better results, visible in the improved match of the line cores for, e.g., lines 8 and 13 (see \fig{fig:muram_inverted_spec_compare}). The retrieved \loggf{} values for lines 8 and 13, listed in \tab{tab:lines}, show deviations that are comparable to errors expected for lines in the NUV \citep[][]{Shchukina13}. 

The results displayed in \fig{fig:muram_atom_pars} also indicate that the spatially-averaged method struggles to recover correct atomic parameters for weaker and blended lines. This is expected and demonstrates the limitation of the spatially-averaged method, where the 1D atmospheric model cannot represent the true temperature stratification in the 3D MHD cube \citep{Han11}. The inversion code mitigates this problem by retrieving an incorrect value of \loggf{} to fit the spatially averaged spectrum better.

It is interesting to note the results achieved for lines 3 and 5, which are rather strong lines but show considerable blending with the wings of line 8. The \globin{} code retrieved sufficiently accurate \loggf{} values for these two lines because of their measurable impact on the $\chi^2$ value. The slight change in their \loggf{} value results in a significant change to the quality of the fit. 

In the presented case, line blending is most prominent in spectral lines 15 and 16 and 17 and 18, where the blending is so strong that both lines appear as a single spectral line. Weak and blended spectral lines show the largest deviation of the retrieved \loggf{inv} from the expected \loggf{0} (\fig{fig:muram_atom_pars}), especially when they are weaker than the rest of the lines in their vicinity.

Comparison with the results from \sref{sec:test_global} shows the advantage of using the \Globin{} instead of the spatially-averaged method, especially for weak and blended spectral lines. Lacking the precise \loggf{} values for weak lines can hinder reliable retrieval of atmospheric parameters from the strong lines with accurate \loggf{} values. This is particularly relevant when applying the spatially-averaged method to the NUV spectral lines where the line blending is more pronounced than in the visible and infrared regions. 

\section{Conclusions}
\label{sec:conclusions}

To infer the stratification of parameters of the solar atmosphere from spectropolarimetric observations, we require precise atomic data of spectral lines (excitation potential of energy levels, elemental abundance, transition probability \loggf{}, central wavelength, collisional broadening, etc.). The atomic parameters are also important in stellar physics for inferring the effective temperature, surface gravity and elemental abundances of different stars \citep[][]{Bigot06,Boeche16,Heiter21}.

In this paper, we have presented a new inversion method (named here as the \Globin{} method) that retrieves atmospheric parameters at every spatial pixel, along with precise values of the atomic parameters even for heavily blended spectral lines. Only a single set of the latter is determined, as they are not allowed to vary from one spatial position to another. This is achieved by requiring the Stokes spectra from all spatial pixels in the given field of view to be fit simultaneously, with the atmospheric parameters being free parameters individually for every pixel, and the atomic parameters being one 'global' set of values for all pixels. This method, implemented in the inversion code \globin{}, follows a similar approach as the method of spatially coupled inversions by \citet{Michiel12}.

We tested the \Globin{} method and compared its results with those from the pixel-by-pixel method for the spectral region 4015\,\AA{}-4017\,\AA{} containing 18, mainly blended, spectral lines. We synthesised line profiles assuming the LTE approximation using the atmospheric models of the umbra, penumbra, granule and intergranular lanes extracted from the MURaM simulation of a sunspot. The \Globin{} method is able to retrieve atmospheric parameters, \loggf{} and \dlam{}, with high precision, whereas the pixel-by-pixel method shows a large deviation in the temperature and \loggf{} from the reference values. This deviation is due to the blending of spectral lines and, to some extent, due to the cross-talk between inversion parameters (mainly between temperature and \loggf{} and \loggf{} and \dlam{}). The pixel-by-pixel method is unable to decouple line contributions to blended profiles, thus inferring erroneous parameter values. 

The inverted line profiles are affected by the goodness of the atmospheric parameterisation used to reproduce the synthetic line profiles. The simple approximation of using a few nodes might not be adequate to represent the highly complex stratification found in MHD atmospheric models of different solar features. As a result of this, we have always contrasted our results for atmospheric parameters to those retrieved in the inversion with fixed atomic parameters (equal to those used to synthesise observed spectra). We found that imposing a coupling in atomic parameters reduces the cross-talk between atmospheric and atomic parameters, retrieving the stratification in atmospheric parameters as best as the chosen node parameterisation can achieve from the given setup.

In producing the synthetic observations, we have ignored any instrumental and observational degradation of spectra, allowing us to compare the performances of the pixel-by-pixel and \Globin{} methods under ideal conditions. Introducing the degradation caused by the photon noise would mask very weak lines, causing the optimisation algorithm to be insensitive to changes in the atomic parameters of those lines. This presents limitations on which lines we can apply our method to infer reliably the atomic parameters.

We showed how the \Globin{} method can aid us in separating contributions of blended line profiles and retrieving accurate \loggf{} values. This result is significant for the analysis of many-line spectropolarimetric observations, such as the NUV spectra expected from the SUSI instrument of the \textsc{Sunrise iii} observatory. The high density of spectral lines in this spectral region results in line blending. Additionally, there are considerable uncertainties in the knowledge of the atomic parameters for many of them. Besides the NUV lines, the \Globin{} method can be used to retrieve the atomic parameters for lines in the visible and infrared wavelengths.

Another issue of analysing the NUV spectra and inferring the atomic parameters of those lines is in properly accounting for the continuum opacity sources. In our study, we have ignored the impact of the continuum opacity contribution from unresolved lines that produce line haze \citep[see e.g.,][]{Greve80,Rutten19} and assumed that we can model the true continuum. Not being able to reproduce the true continuum in the observed spectrum will be compensated by adjusting the retrieved temperature stratification and \loggf{} of lines. Currently, the opacity fudge coefficients \citep[e.g.,][]{Bruls92} are used to enhance the calculated continuum opacity and reproduce the observed continuum.

Furthermore, we inferred the \loggf{} values for the spectral lines in the same spectral range by fitting the spatially averaged synthetic spectrum computed from the quiet-sun MURaM atmospheric model. The 1D atmospheric model is inferred from the spatially averaged spectrum, assuming that atomic parameters are known. We then use this atmospheric model and repeat the inversion to infer only the \loggf{} value of each line. The initial \loggf{} values are displaced from the exact values by Gaussian randomization. The inferred results for strong and unblended lines match the exact values very well, while significant scattering is obtained for weaker and more blended spectral lines. This finding is in line with the use of the spatially averaged spectrum of the Sun to derive the \loggf{} values assuming a representative 1D atmospheric model (the spatially-averaged method). The spatially-averaged method is generally restricted to isolated and strong spectral lines in the visible and infrared wavelengths \citep[see for example][]{Borrero03}.

In all the inversions presented in this paper, we have assumed that the used line list is complete (all lines are known and accounted for during synthesis and inversion). This is not always the case for observed spectra. Disregarding one of the lines will result in an erroneous \loggf{} value of other lines sufficiently blended with the disregarded line. The impact of missing lines on the inferred atmospheric stratification and atomic parameters is currently under investigation and will be reported in a separate publication. Additionally, the accuracy of the inferred \loggf{} is also impacted by the poor knowledge of other atomic line parameters, such as the line central wavelength and the excitation potential of the lower energy level. We still rely heavily on laboratory measurements to provide precise values of these parameters. The poor knowledge of atomic parameters does not refer only to the line of interest, but also to other lines that are part of the blend.

We assumed in all inversions that lines are properly modelled in the LTE approximation, thus removing any discrepancies between the synthetic and inverted line profiles that might arise from different modelling assumptions. NLTE effects, important for the formation of many spectral lines in the considered spectral region, can be modelled with \globin{} and pose no additional problem in the inference of the above-mentioned atomic parameters. It is not yet clear how big of an impact the LTE approximation has on the inference of the \loggf{} parameter for lines that should be treated in NLTE. We plan to investigate this for the widely used line pair of Fe\,I at 6301.5\,\AA{} and 6302.5\,\AA{}, which is shown to be sensitive to NLTE effects \citep{Smitha21}.

We intend to extend the \Globin{} method and add elemental abundance as another inversion parameter that is considered as a global parameter having a unique value across the observed field of view. Inverting a spectral region that contains many spectral lines of a single atomic element will impose additional constraints on the inference of the element abundance. We believe it is possible to disentangle the contribution of the elemental abundance and \loggf{} to the line opacity by having at least one line in the observed spectral region with reliable \loggf{}. This will bring a completely new and independent way of determining the solar elemental abundances, which are currently determined mainly using 3D hydrodynamical models of the solar atmosphere \citep[see for e.g.,][]{Asplund09} that already use prescribed elemental abundances.

\begin{acknowledgements}
       D.V. is founded by International Max Planck Research School (IMPRS) for Solar System Science at the University of Göttingen. We thank M. Rempel for kindly providing the MHD cubes. This project has received funding from the European Research Council (ERC) under the European Union's Horizon 2020 research and innovation programme (grant agreement No. 101097844 — project WINSUN). This work has made use of the VALD database, operated at Uppsala University, the Institute of Astronomy RAS in Moscow, and the University of Vienna.
\end{acknowledgements}

\bibliography{references.bib}

\onecolumn

\begin{appendix}

\section{Pixel-by-pixel method}
\label{sec:pxl-by-pxl}

The inversion of spectropolarimetric observations is a non-linear optimization problem. A regularly used optimization scheme is the Levenberg-Marquardt algorithm \citep[][henceforth LM]{Levenberg44,Marquardt63}, which is a combination of the gradient descent and the Gauss-Newton methods for minimizing a merit function. 

An important part of every optimization procedure is to define a merit function to be minimized through the iterative correction of free parameters. Spectropolarimetric inversion algorithms generally use $\chi^2$ as a merit function, which is defined as \citep[][]{Iniesta03}:
\begin{equation}\label{eq:chi2}
    \chi^2(\mathbf{p}) = \frac{1}{N - n}\sum_{i=1}^N \frac{w_{i}^2}{\sigma_i^2} \cdot \left( \mathbf{O}_{i} - \mathbf{S}(\mathbf{p})_{i}  \right)^2,
\end{equation}
where the index $i$ goes over each wavelength point for all four Stokes components ($N=4N_\lambda$, $N_\lambda$ being the number of wavelength points), and $n$ the number of free parameters. The factor $w_{i}$ is the weight given to each wavelength and $\sigma_{i}$ is the corresponding noise at this wavelength. $\mathbf{O}_{i}$ denotes the observed Stokes vector and $\mathbf{S}(\mathbf{p})_{i}$ denotes the synthetic Stokes vector, with $\mathbf{p}$ representing a vector of free parameters (temperature $T$, line-of-sight velocity $v_\mathrm{LOS}$, micro-turbulent velocity $v_\mathrm{mic}$, magnetic field strength $B$, magnetic field inclination $\theta$, magnetic field azimuth $\phi$, $\log(gf)$, $\Delta \lambda$). The Stokes vector is defined as $(I,Q,U,V)$.

The LM algorithm uses the first derivative of a model function to minimize the merit function that defines a hypersurface in the inversion parameter space. The $\chi^2$ hypersurface may have many local minima and usually only one global minimum corresponding to the best inversion parameters set. However, some local minima can be almost as deep as the global minimum because of the cross-talk between different inversion parameters. The LM algorithm efficiently finds a minimum of the merit hypersurface but does not guarantee that the found minimum corresponds to the global one. By adjusting the weights $w_{i}$ of different wavelengths in the spectrum, it is possible to change the shape of the merit hypersurface. Such an adjustment can produce a more pronounced global minimum and increase the efficiency of finding the global minimum of the merit function. Finding the best fit to the observed spectrum depends on how close the initial solution is to the optimal one, the complexity of the model function and the employed weights.

In the LM algorithm, an initial guess for the inversion parameters is assumed to be close to the global minimum. Expanding the merit function around the global minimum in parameter space using a second-order Taylor polynomial yields \citep{Press07}:
\begin{equation}
    \chi^2(\mathbf{p}) = \chi^2(\mathbf{p_j}) + \Delta \mathbf{p_j^\mathrm{T}}\cdot\bigtriangledown \chi^2(\mathbf{p_j}) + \frac{1}{2} \Delta \mathbf{p_j^\mathrm{T}}\cdot\mathcal{H}\cdot\Delta \mathbf{p_j},
    \label{eq:taylor_poly}
\end{equation}
where $\Delta\mathbf{p_j} = \mathbf{p} - \mathbf{p_j}$ is a correction for the parameter vector $\mathbf{p_j}$ in the $j$--th iteration, $\mathcal{H}$ is the Hessian matrix of the system, and $\mathrm{T}$ indicates the matrix transpose. Based on our assumption being at the minimum, we expect the gradient of $\chi^2$ to be zero, $\bigtriangledown\chi^2(\mathbf{p}) = 0$. Thus, taking a gradient of \eq{eq:taylor_poly}, we obtain:
\begin{equation}
    0 = \bigtriangledown\chi^2(\mathbf{p_j}) + \mathcal{H}\cdot\Delta\mathbf{p_j},
    \label{eq:gradient_LM}
\end{equation}
where $\frac{\mathrm{d}\Delta\mathbf{p_j}}{\mathrm{d}\mathbf{p}}=1$. In this derivation, we have used the identity:
\begin{equation*}
    \frac{\mathrm{d}\left(\Delta\mathbf{p_j^\mathrm{T}}\cdot\mathcal{H}\cdot\Delta \mathbf{p_j}\right)}{\mathrm{d}\mathbf{p}} = \left(\mathcal{H} + \mathcal{H}^\mathrm{T} \right)\cdot\Delta\mathbf{p_j} = 2\mathcal{H}\cdot\Delta\mathbf{p_j},
\end{equation*}
where we applied the symmetry property of the Hessian matrix, $\mathcal{H}\equiv\mathcal{H}^\mathrm{T}$. Solving \eq{eq:gradient_LM} gives the correction $\Delta \mathbf{p_j}$ for the initial parameter vector that minimizes $\chi^2$. Here we have implicitly assumed that the correction of the inversion parameters produces a linear change in the model, which produces quadratic change in the $\chi^2$. Precisely this linearisation of the non-linear model requires an iterative correction of inversion parameters.

The gradient of the merit function from \eq{eq:gradient_LM} is:
\begin{equation}
    -\frac{\partial\chi^2}{\partial p_k} = \frac{2}{N-n}\sum_{i=1}^N \frac{w_{i}^2}{\sigma_i^2} \cdot \left( \mathbf{O}_{i} - \mathbf{S}_i(\mathbf{p})  \right)\cdot\frac{\partial\mathbf{S}_i}{\partial p_k},
    \label{eq:chi2_grad}
\end{equation}
where the term $\frac{\mathrm{d}\mathbf{S}_i}{\mathrm{d}p_k}$ is the response function of the Stokes spectrum to the $k$--th inversion parameter at wavelength $i$ \citep[][]{Iniesta03}. With further differentiation of \eq{eq:chi2_grad} with respect to parameter $p_l$, we obtain the Hessian matrix elements:
\begin{equation}
    \mathcal{H}_{l,k} = \frac{\partial\chi^2}{\partial p_l \partial p_k} = \frac{2}{N-n}\sum_{i=1}^N \frac{w_{i}^2}{\sigma_i^2} \cdot \left[\frac{\partial\mathbf{S}_i}{\partial p_k}\cdot\frac{\partial\mathbf{S}_i}{\partial p_l} - \left( \mathbf{O}_{i} - \mathbf{S}_i(\mathbf{p})  \right)\cdot\frac{\partial^2\mathbf{S}_i}{\partial p_l \partial p_k}\right].
    \label{eq:chi2_hessian}
\end{equation}
The second term under square brackets in \eq{eq:chi2_hessian} is usually disregarded in LM inversions \citep{Iniesta03,STIC}. This is reasonable since it was assumed that the initial solution was close to the minimum. Therefore, $\mathbf{O} - \mathbf{S(\mathbf{p_j})}$ should be zero and the second-order derivatives will not influence the proposed parameter steps. Disregarding second-order derivatives ensures that the Hessian matrix is positive definite, and the step in parameter space leads to the minimisation of $\chi^2$. In some occasions, the second-order derivatives can even lead to corrupted behaviour \citep{Press07}. 

Substituting \eq{eq:chi2_grad} and \eq{eq:chi2_hessian} into \eq{eq:gradient_LM}, and identifying new terms, we obtain:
\begin{equation}
    \mathcal{H}\cdot\Delta \mathbf{p_j} = \mathcal{J}^\mathrm{T}\cdot\mathbf{\Delta},
    \label{eq:LM}
\end{equation}
where $\mathbf{\Delta}_i = \sqrt{\frac{2}{N-n}}\cdot\frac{w_i}{\sigma_i}\cdot\left( \mathbf{O}_i - \mathbf{S}_i(\mathbf{p_j}) \right)$ and $\mathcal{J}$ is the Jacobian matrix of the system whose elements are given as:
\begin{equation*}
    \mathcal{J}_{i,k} = \sqrt{\frac{2}{N-n}}\frac{w_i}{\sigma_i}\frac{\partial\mathbf{S}_i}{\partial p_k}.
\end{equation*}
The linearisation of the Hessian matrix allows us to express it through the Jacobian matrix as $\mathcal{H}=\mathcal{J}^\mathrm{T}\mathcal{J}$.

Fitting non-linear models to observed data can lead to poor parameter corrections. To address this issue, the diagonal elements of the Hessian matrix are multiplied with a factor $\lambda_\mathrm{M}$ and added to the Hessian, yielding $\mathcal{H} = \mathcal{J}^\mathrm{T}\mathcal{J} + \lambda_\mathrm{M}\cdot \mathrm{diag}(\mathcal{J}^\mathrm{T}\mathcal{J})$. This factor is known as the Marquardt parameter and regulates the magnitude of the parameters correction (i.e., $\Delta p_k \propto 1/\lambda_\mathrm{M}$). For fast convergence, the Marquardt parameter should be small enough (e.g., $\lambda_\mathrm{M} < 10^{-2}$) but not too small to overstep the global minimum. 
 
\section{the \Globin{} method -- algorithm layout}
\label{sec:global}

In the pixel-by-pixel method, the Jacobian and the Hessian matrices are constructed for every pixel where eq.\,\ref{eq:LM}  provides a parameter correction independently from other pixels. In the \Globin{} method, a global parameter has a uniform value in all pixels whose retrieval requires a coupling between the pixels. This coupling is introduced by requiring a simultaneous match in the observed and inverted Stokes spectra from all the pixels. In the \Globin{} method, inversion parameters are obtained by minimizing the sum of $\chi^2$ from all the pixels:
\begin{equation}\label{eq:chi2global}
    \chi_\mathrm{global}^2(\mathbf{p}) = \frac{1}{N-N_\mathrm{p}}\sum_{a=1}^{N_\mathrm{atm}}\sum_{i=1}^N \frac{w_{i}^2}{\sigma_i^2} \cdot \left( \mathbf{O}_{i,a} - \mathbf{S}(\mathbf{p})_{i,a}  \right)^2,
\end{equation}
where the index $a$ goes over every atmosphere in the considered field-of-view containing the total number of atmospheres $N_\mathrm{atm}$, and $N_\mathrm{p}$ is the total number of free parameters summed over all pixels. Here, the parameter vector $\mathbf{p}$ contains parameters from every pixel in the field-of-view $\mathbf{p} = (\mathbf{p}_1, \mathbf{p}_2, \ldots, \mathbf{p}_\mathrm{N_{atm}})$.

In the \Globin{} method, the functional dependence of the $\chi^2$ function does not change except for the addition of a summation which runs over all pixels (atmospheres). Therefore, following the same procedure as in the pixel-by-pixel inversion, we derive an equation for the correction of the parameters for the \Globin{} inversions:
\begin{equation}
    \mathcal{H_\mathrm{global}} \Delta \mathbf{p_j} = \mathcal{J}_\mathrm{global}^\mathrm{T} \left( \mathbf{O} - \mathbf{S}(\mathbf{p_j})  \right),
\end{equation}
where $\mathcal{H}_\mathrm{global}$ and $\mathcal{J}_\mathrm{global}$ are the global Hessian and Jacobian matrices of the system, respectively. The global Hessian and Jacobian matrices are connected in the same manner as in the pixel-by-pixel method. Therefore, to explain how the \Globin{} method works, it would be sufficient to derive the global Jacobian matrix.

In the pixel-by-pixel method, column $k$ in the Jacobian matrix contains the response function of the Stokes vector $\mathbf{R}_k$ to the $k$-th inversion parameter, where $\mathbf{R}_k$ is a vector of length $4 \times N_\lambda$. In case of $n$ inversion parameters, the Jacobian matrix for a single pixel has a dimension $(4 \times N_\lambda, n)$ and can be written as $\mathcal{J} = (\mathbf{R}_1, \mathbf{R}_2, \ldots, \mathbf{R}_n)$.

The global Jacobian matrix of the system is constructed by placing all the single pixel Jacobian matrices on the diagonal:

\begin{equation}
    \mathcal{J}_\mathrm{global} = \left( 
        \begin{array}{ccccccc}
            \mathcal{J}_{1} & 0 & \cdot & \cdot & \cdot & & 0 \\
            0 & \mathcal{J}_{2} \\
            \cdot & & \cdot & & & & \cdot \\
            \cdot & & & \cdot & & & \cdot \\
            \cdot & & & & \cdot & & \cdot \\
             & & & & & \mathcal{J}_{N_\mathrm{atm-1}} & 0 \\
            0 & & \cdot & \cdot & \cdot & 0 & \mathcal{J}_{N_\mathrm{atm}}
        \end{array}
        \right).
\end{equation}

This matrix is block-diagonal and ensures an uncoupled inversion of parameters for each pixel. The transposed form of a block-diagonal matrix is a block-diagonal matrix of transposed sub-matrices resulting in a block-diagonal global Hessian matrix. The global Jacobian matrix corresponds to the system with $n \times N_\mathrm{atm}$ number of free parameters and has a dimension $(4 \times N_\lambda \times N_\mathrm{atm},\ n \times N_\mathrm{atm})$.

The block-diagonal form of the global Jacobian matrix has to be disrupted to achieve a coupled inversion. Therefore, let us assume that out of $n$ parameters for each pixel we have $l$ local (atmospheric) and $g$ global (atomic) parameters ($n=l+g$). The atmospheric parameters vary between pixels due to the different physical structures of each pixel while atomic parameters are the same for every pixel. In the \Globin{} method, the total number of free parameters is therefore $N_\mathrm{atm}\times l + g$, whereas in the pixel-by-pixel method this number is $N_\mathrm{atm}\times(l+g)$. Additionally, we assume that the inversion parameters in the vector $\mathbf{p}$ are ordered from atmospheric to atomic.

Substituting the Jacobian sub-matrices with the response functions we get: 

\begin{equation}
\mathcal{J}_\mathrm{global} = \left( 
        \begin{array}{ccccccc}
            \mathbf{R}_1^1\ldots\mathbf{R}_l^1\;\mathbf{R}_{l+1}^1\ldots\mathbf{R}_{l+g}^1 & 0 & \cdot & \cdot & \cdot & & 0 \\
            0 & \mathbf{R}_1^2\ldots\mathbf{R}_l^2\;\mathbf{R}_{l+1}^2\ldots\mathbf{R}_{l+g}^2 & & & & & \cdot \\
            \cdot & & \cdot & & & & \cdot \\
            \cdot & & & \cdot & & & \cdot \\
            \cdot & & & & \cdot & & 0 \\
            0 & & \cdot & \cdot & \cdot & & \mathbf{R}_1^{\mathrm{N_{atm}}}\ldots\mathbf{R}_l^{\mathrm{N_{atm}}}\;\mathbf{R}_{l+1}^{\mathrm{N_{atm}}}\ldots\mathbf{R}_{l+g}^\mathrm{N_{atm}}
        \end{array}
        \right),
\end{equation}
where the upper index in the response function corresponds to the pixel number.

To have the global inversion of atomic parameters, we have to reorder the response functions in this matrix. For a single global parameter we take its response functions for all pixels and form a single column. Consequently, for the $k$-th parameter we have $\left(\mathbf{R}_k^1, \mathbf{R}_k^2,..., \mathbf{R}_k^\mathrm{N_{atm}}\right)^\mathrm{T}$. Further, we repeat this process for each global parameter we have in the inversion. This results in a total of $g$ columns that contain only the response functions to the global parameters. Then, we move these columns to the right side of the matrix. The rest of the matrix now contains the response functions to the local parameters only. These response functions form a sub-matrix which has block-diagonal form. Therefore, the global Jacobian matrix after the reordering in the response functions is:

\begin{equation}
\mathcal{J}_\mathrm{global} = \left( 
        \begin{array}{ccccclcc}
            \mathbf{R}_1^1\ldots\mathbf{R}_l^1 & 0 & \cdot & \cdot & \cdot & & 0 & \mathbf{R}_{l+1}^1\ldots\mathbf{R}_{l+g}^1 \\
            0 & \mathbf{R}_1^2\ldots\mathbf{R}_l^2 & & & & & \cdot & \mathbf{R}_{l+1}^2\ldots\mathbf{R}_{l+g}^2 \\
            \cdot & & \cdot & & & & \cdot & \cdot \\
            \cdot & & & \cdot & & & \cdot & \cdot \\
            \cdot & & & & \cdot & & 0 & \cdot \\
            0 & & \cdot & \cdot & \cdot & & \mathbf{R}_1^\mathrm{N_{atm}}\ldots\mathbf{R}_l^{\mathrm{N_{atm}}}&\mathbf{R}_{l+1}^\mathrm{N_{atm}}\ldots\mathbf{R}_{l+g}^\mathrm{N_{atm}}
        \end{array}
        \right).
\end{equation}

The block-diagonal form of the global Jacobian matrix is retained only for the local parameters and the coupling is introduced only for the inversion of global parameters. Corrections of global parameters are determined from spectrum differences in all pixels, while corrections for local parameters are determined from the spectral difference in a given pixel. This reordering keeps the pixel-by-pixel inversion of atmospheric parameters and introduces spatial coupling of, e.g., atomic parameters.

The global Jacobian matrix with coupling in atomic parameters has dimension $(4 \times N_\lambda \times N_\mathrm{atm},\ N_\mathrm{atm}\times l + g)$ which is lower in comparison to a global Jacobian matrix for uncoupled inversion due to the grouping of the response functions of the global parameters. This coupling lowers the number of free parameters in the inversion, resulting in a $\chi^2$ hypersurface that should produce fewer local minima. In the \Globin{} method, we invert the same number of data points as in the pixel-by-pixel method with fewer parameters. This adds more constraints to the inversion parameters and aids the inversion algorithm in finding the global minimum.

In the \Globin{} method, a single value of the Marquardt parameter controls the step sizes for all parameters, whereas in the pixel-by-pixel method, each pixel has its own Marquardt parameter. This difference has consequences in effectively locating the global $\chi^2$ minimum. A similar set of equations are solved in the PSF coupled inversion algorithm of \cite{Michiel12}. In that paper, the author argues that to effectively achieve convergence to the global minimum, the inversion should be run for a small number of iterations ($\sim$10) after which the parameters are perturbed and used as initial values for the following inversion run. This way, the LM algorithm is kicked from any local minimum it may have strayed into, to locating the global minimum.

\section{\globin{}}
\label{sec:globin}

Current spectropolarimetric inversion codes like SIR \citep{SIR}, SPINOR \citep{SamiThesis,SPINOR} and NICOLE \citep{NICOLE} can invert both atmospheric and atomic parameters such as \loggf{}, the central wavelength of line, elemental abundance and line broadening parameters. However, these parameters are allowed to freely vary at every pixel in these codes. Here, we will describe in detail the implemented functionality of the \globin{} code in which we have implemented the new method for retrieving atomic parameters.

Depth variations of physical parameters in an atmospheric model are commonly assumed to be a function of the optical depth, calculated at a reference wavelength of $5000$\,\AA{}. To properly synthesize different spectral lines (e.g. NUV lines), we require the stratification of atmospheric models from photospheric up to chromospheric layers to be specified on a fine grid. Inverting observed spectra by iterative adjustment of all atmospheric parameters at each depth is very complicated due to the large number of free parameters. A feasible inversion of atmospheric structure therefore requires an approximation of the stratification of atmospheric parameters. 

One kind of approximation is to parameterize the atmospheric stratification using a few points at properly chosen depths and interpolate between them to obtain values on a finer grid as needed for the spectral synthesis \citep[][]{SIR, Iniesta03}. These points are called nodes and are placed over the range over which the line and its nearby continuum respond to changes in the atmosphere. In \globin{}, nodes contain values of parameterized atmospheric parameters. Node positions need to be specified on an optical depth scale and can be chosen for each parameter independently. The interpolation is done either using non-overshooting Bezier interpolation polynomials of 2nd or 3rd order \citep[][]{Jaime13}, or cubic splines under tension. The polynomial approximation, interpolation degree and the spline tension are chosen by the user. The values of the parameters from the highest node to the top boundary of the atmosphere are linearly extrapolated. The same procedure is also applied from the deepest node to the bottom boundary. For the temperature, we linearly extrapolate using the temperature gradient from the FAL-C model \citep[][]{FALC} for the lowest node \citep{SNAPI}. The temperature extrapolation to the higher layers has a lower limit of $2800$\,K in the highest point in the atmosphere. The same limit is also imposed on all temperature nodes. For lower temperatures, most hydrogen atoms form H$^2$ molecule, significantly reducing the population of neutral hydrogen atoms. This affects the continuum opacity contributions, such as from H$^{-}$ ion, which produces a significantly different continuum level compared to the observed one.

The most time-consuming part of any inversion is the computation of the response functions. In the LTE approximation, the response functions are computed analytically at the same time when the radiative transfer equation is solved \citep{SIR}. In \globin{}, we opted for a straight forward brute-force method and compute response functions using a numerical central difference scheme \citep[see e.g.][]{QuinteroNoda16}. From the initial value of the inversion parameter, we make a positive perturbation by a small amount $\delta p_k$ (see Table\,\ref{tab:perturbations}), interpolate to obtain a refined stratification, and compute the resulting Stokes spectrum $\mathbf{S^+}$. Next, starting from the same initial value, we repeat the process with a negative perturbation of the same magnitude and compute the corresponding Stokes spectrum $\mathbf{S^-}$. The response function of a given parameter is computed as $\frac{\mathbf{S^+} - \mathbf{S^-}}{2\delta p_k}$. This numerical approach to compute the response function allows the application of the method to arbitrarily complex atmospheric stratifications.

The perturbation values given in Table\,\ref{tab:perturbations} are chosen to be small enough to approximate the response functions as a first-order perturbation of the spectrum, but large enough to produce differences in the spectra that are significantly larger than any numerical uncertainty. In the case of much larger perturbations of the physical parameters, we would reach a non-linear perturbation of the spectrum due to the non-linearity of the radiative transfer equation.

\begin{table}
    \centering
    \caption{The parameter perturbations $\delta p$ used for computing numerical response functions.}
    \begin{tabular}{c c}
    \toprule[1.5pt]  
        parameter & perturbation \\ 
        \midrule[1.5pt] 
        $T$ & 1 K \\
        $v_\mathrm{LOS}$ & 1 m/s \\
        $v_\mathrm{mic}$ & 1 m/s \\
        $B$ & 1 G \\
        $\gamma$ & 0.01 rad \\
        $\phi$ & 0.01 rad \\
        $\log(gf)$ & 0.001 \\
        $\Delta \lambda$ & 1 m\AA{} \\
    \bottomrule[1.5pt] 
    \end{tabular}
    \label{tab:perturbations}
\end{table}

The values of the response functions for different physical parameters span several orders of magnitude. The origin of this lies in the sensitivity of the Stokes spectra to perturbations of particular physical parameters and the choice of units for these parameters. It leads to unequal scaling between parameters, which influences the convergence properties of the LM algorithm. To improve the convergence of the inversion, \cite{Marquardt63} suggested scaling the response function of each $k$--th parameter with:
\begin{equation}
    s_k = \sqrt{\sum_{i=1}^N J_{i,k}^2},
\end{equation}
where the index $i$ goes over all wavelengths for all four Stokes components. Dividing the computed response function with this scaling parameter yields a dimensionless response function. The Hessian matrix then becomes a matrix of parameter correlation coefficients (diagonal elements are equal to $1+\lambda_\mathrm{M}$). The parameter correction $\Delta p_k$ can be converted back to proper units by multiplying it with the $s_k$.

During each iteration of the inversion process, the atmospheric model is assumed to be in hydrostatic equilibrium, so that the temperature alone is sufficient to derive gas and electron pressures assuming the ideal gas law. From these, we obtain the populations of hydrogen in the LTE approximation, necessary to compute spectra using the RH code. 

\end{appendix}

\end{document}